# Automated Data-Driven Discovery of Material Models Based on Symbolic Regression: A Case Study on Human Brain Cortex

Jixin Hou[1], Xianyan Chen[2], Taotao Wu[3], Ellen Kuhl[4], Xianqiao Wang[1]*

[1]School of Environmental, Civil, Agricultural and Mechanical Engineering, College of Engineering, University of Georgia, Athens, GA, 30602, USA

[2]Department of Epidemiology and Biostatistics, College of Public Health, University of Georgia, Athens, GA 30602, USA

[3]School of Chemical, Materials, and Biomedical Engineering, College of Engineering, University of Georgia, Athens, GA, 30602, USA

[4]Department of Mechanical Engineering, Stanford University, Stanford, CA 94305, USA

*Corresponding Author: xqwang@uga.edu

## Abstract

We introduce a data-driven framework to automatically identify interpretable and physically meaningful hyperelastic constitutive models from sparse data. Leveraging symbolic regression, an algorithm based on genetic programming, our approach generates elegant hyperelastic models that achieve accurate data fitting through parsimonious mathematic formulae, while strictly adhering to hyperelasticity constraints such as polyconvexity. Our investigation spans three distinct hyperelastic models—invariant-based, principal stretch-based, and normal strain-based—and highlights the versatility of symbolic regression. We validate our new approach using synthetic data from five classic hyperelastic models and experimental data from the human brain to demonstrate algorithmic efficacy. Our results suggest that our symbolic regression robustly discovers accurate models with succinct mathematic expressions in invariant-based, stretch-based, and strain-based scenarios. Strikingly, the strain-based model exhibits superior accuracy, while both stretch- and strain-based models effectively capture the nonlinearity and tension-compression asymmetry inherent to human brain tissue. Polyconvexity examinations affirm the rigor of convexity within the training regime and demonstrate excellent extrapolation capabilities beyond this regime for all three models. However, the stretch-based models raise concerns regarding potential convexity loss under large deformations. Finally, robustness tests on noise-embedded data underscore the reliability of our symbolic regression algorithms. Our study confirms the applicability and accuracy of symbolic regression in the automated discovery of hyperelastic models for the human brain and gives rise to a wide variety of applications in other soft matter systems.

# 1. Introduction

Solving a solid mechanical problem with provided information on displacements or forces requires adherence to three fundamental principles: the kinematic equations, delineating the relations between displacements and strains; the balance principles, establishing the connection between interior stresses and external tractions or body forces; and the constitutive equations, serving as the intermediary bridge between strains and stresses [1, 2]. As intrinsic material properties commonly characterized through experiments, constitutive relationships articulate mechanical behaviors within specific configurations, playing indispensable roles in engineering analyses and *in silico* simulations [3-7]. In the general context, we formulate constitutive relationships deploying various formats to capture distinct material responses, including elasticity, hyperelasticity, viscoelasticity, or plasticity. A more concise approach involves employing Onsager's thermodynamic framework [8, 9], where a scalar thermodynamic potential proves sufficient for fully characterizing the material behaviors of diverse materials. Specifically, this individual potential comprises two independent counterparts: the reversible Helmholtz free energy potential and the irreversible dissipation potential [10]. In cases with no dissipative effects like plasticity or viscosity, the Helmholtz free energy alone is capable of characterizing the mechanical response, as in our current study.

An elegant constitutive model should not only exhibit prominent accuracy in characterizing material behaviors but also preserve precision in dealing with perturbations or noise from experimental measurements. To uphold these qualities, constitutive models must rigorously adhere to physical laws such as thermodynamic consistency and possess well-posed mathematical representations [11]. Furthermore, these models should be exquisitely crafted to meet the criteria of objectivity, stability, and, if desired, material symmetry [12, 13]. Material stability is crucial to ensure the existence and uniqueness of solutions for boundary-value problems [14, 15]. Its mathematical representation equates to the convexity condition, which is closely linked to the positive definiteness of the second-order Hessian matrix of the free energy functions [16]. However, imposing convexity is often overly restrictive from a physical perspective and proves challenging in practical mathematical applications. An alternative condition is the polyconvexity condition [17], which serves as a sufficient condition for material stability and demonstrates remarkable superiority in mathematical representations [18]. When coupled with the growth condition that defines unbounded energy for infinite deformations, the polyconvexity condition translates into the necessary and sufficient condition for stability, and ensures the existence of the solution [19]. Hence, to ensure accurate and precise performance, during the design of constitutive models, it is imperative to enforce the above physical constraints.

Constitutive relations are commonly characterized through experiments that typically record data in the form of displacement-force pairs or strain-stress pairs as their derivatives. The conventional strategy involves directly calibrating this data using regression algorithms, such as the least square method. However, in these methods, a pre-established material model must be defined prior to calibration. Therefore, the efficacy of calibration is heavily reliant on the initial model selection, which is significantly influenced and biased by individual experiences. Typically, massive efforts are invested in iteratively seeking an appropriate material model, resulting in a tedious and laborious calibration procedure [10]. More promising and automated approaches are the data-driven techniques, especially neural networks, which are primarily built on machine learning backbones and have emerged as versatile tools for facilitating the discovery of constitutive models. Classic neural networks, such as Feed-Forward Neural Network (FFNN), adopt fully connected structures and are commonly trained with strain input and stress output. While these networks can accurately interpolate experimental data, they may struggle with overfitting and fail to extrapolate well outside the training regime [19, 20]. To address these limitations, Physics Informed Neural Networks (PINN) are introduced, with particular considerations of physical laws [21]. This approach mainly operates in two ways: i) customizing loss functions by introducing additional terms to penalize the violation of physical laws [22-24]; and ii) crafting network architectures in accordance with the physical validity constraints [25, 26]. Representative examples are the Constitutive Artificial Neural Network (CANN) [27], Input Convex Neural Network (ICNN) [13], and the Neural Ordinary Differential Equations (NODE) [28]. Comprehensive explanations and benchmark tests of the CANN, ICNN, and NODE models are available in a recent review [29]. Additional efforts include the Gaussian process [30], spline approximation [31], and probability inference [32]. Moreover, unsupervised investigations into constitutive models, such as the Efficient Unsupervised Constitutive Law Identification and Discovery (EUCLID) model, have also been explored [10].

Data-driven methods based on machine learning show promising applications in the automated discovery of constitutive models. However, most of the previously discussed methods share the following two weaknesses: i) *Black Box Nature*: The model discovery process often acts as a "black box", making it impossible to express predictions in explicit mathematic formulas. This lack of transparency largely restricts the interpretability and portability of the predicted model. For example, if we intend to integrate the predicted model in computational simulations, we will have to construct a configurational interface between the neural network model and computational software to maintain spontaneous interactions. This contrasts with the simplicity of embedding a single mathematical model directly into computational algorithms. ii) *Limited Functional Space*: The available functional space for model selection is often constrained. For instance, the ultimate hyperelastic models, in CANN, are confined to a functional set comprising only 12 exquisitely designed terms [33]. In EUCLID, the identification of the material type and calibration of model parameters are simultaneously achieved by determining finite material parameters within a generalized material library [10]. This limitation could potentially lead to challenges in navigating multiple local optima, thereby missing the global optimal constitutive model.

Symbolic regression, an alternative data-driven approach widely used in scientific research [34-38], differs from the machine learning-based methods by employing genetic programming algorithms. This approach can automatically decipher mathematic information from data without the need for specific a priori knowledge about the investigated systems, thereby significantly enhancing interpretability [39]. Operating on tree structures, symbolic regression iteratively searches for candidate algebraic models, constituted by input variables, mathematic operators, and constants, that gradually match the provided data in an evolutional manner. Theoretically, the functional space in symbolic regression can be considered infinite. The utilization of symbolic regression in constitutive modeling has gained popularity in recent years. For example, sparse symbolic regression has been used to identify algebraic stress models from high-fidelity simulation data, with the predicted models exhibiting significant superiority over traditional turbulent models [40]. Symbolic regression was applied in model characterization and parameter calibration, specifically in the plasticity regime [41]. Recent studies leverage the integration of PINNs with symbolic regression, and successfully discovered novel reaction-diffusion models for the spatio-temporal diffusion of misfolded tau proteins in Alzheimer's disease [23].

In the realm of hyperelasticity, however, few endeavors have explored the use of symbolic regression. Limited studies involve a first attempt to characterize the multi-axial loading behaviors of vulcanized rubber [42, 43], and recent work on an intriguing cooperation of symbolic regression with a neural network model [44]. Here the neural network is employed to facilitate the differentiation operation and enforce physical admissibility laws, while the symbolic regression serves as a mathematical toolbox to generate algebraic formulas, thereby improving the model's interpretability. Though achieving satisfactory prediction for vulcanized rubber and particle-reinforced composites, the multiple simultaneous interactions between symbolic regression and neural networks inevitably increase the computational cost, especially when the anticipated model incorporates a complicated format. Therefore, it remains an unanswered question whether and how we can discover hyperelastic models, with rigorous adherence to physical constraints, solely within the framework of symbolic regression.

In this study, we aim to explore the capabilities of symbolic regression in automatedly discovering hyperelastic models that rigorously comply with physical requirements, such as the polyconvexity condition. To achieve this objective, we investigate three distinct hyperelastic scenarios, invariant-based, principal stretch-based, and normal strain-based hyperelastic models. These investigations are based on multi-mode experimental data from the human brain cortex. In accordance with physical constraints, we meticulously design the model structures, especially focusing on the objective functions. The structure of this paper is organized as follows: First, we introduce the theory of constitutive modeling, with particular emphasis on physical constraints, and the symbolic regression algorithms, along with implementation details in Section 2. Following validation against multiple synthetic datasets, the approaches are implemented on experimental data to discover hyperelastic models for the human brain cortex, with results and discussion presented in Section 3. Finally, we conclude our findings and outline potential directions for future explorations in Section 4.

## 2. Theoretical Method and Symbolic Regression Algorithm

In this section, we revisit the fundamental theorem of continuum mechanics and delineate crucial constraints essential for ensuring the physical admissibility of strain energy function derived through symbolic regression. First, we briefly review the descriptions pertaining to the kinematic equations, balance equations, and constitutive equations within the framework of continuum mechanics. Then, we delve into the critical conditions necessary to acknowledge the physical constraints of strain energy function, especially the condition of polyconvexity. Once these foundational aspects are established, we proceed to predict the general form of the strain energy function for human brain tissue based on experimental data using symbolic regression, the algorithm and implementation details of which will be introduced subsequently.

## 2.1. Constitutive modeling

In the context of continuum mechanics, the kinematics of a continuum body can be described by a one-to-one mapping denoted as $x = \varphi(X)$, where a material particle initially positioned at $X$ in the reference configuration $\mathcal{B}_0$ is carried to its new position $x$ in the current configuration $\mathcal{B}_t$. Quantitatively, we employ the deformation gradient $F = \nabla_X \varphi$ to quantify the mapping of the line element from reference to current configuration, and the Jacobian $J = \det F$ to describe the associated volume alternation. The deformation gradient tensor $F$ has the spectral representation as $F = \nabla_X \varphi = \sum_{i=1}^{3} \lambda_i \, \boldsymbol{n}_i \otimes \boldsymbol{N}_i$, where $\lambda_i$ are the principal stretches, $\boldsymbol{n}_i$ and $\boldsymbol{N}_i$ correspond to the principal directions in the current and reference configurations. For any physically admissible deformation, the Jacobian $J$ consistently remains positive. $J < 1$ indicates a contraction deformation, while $J > 1$ signifies dilation. Left multiplying $F$ by its transpose $F^T$ yields the right Cauchy-Green deformation tensor $C = F^T F$, which is favorable for its numerical convenience owing to its symmetric and positive-definite attributes. The right Cauchy-Green deformation tensor possesses three complete and irreducible principal scalar invariants,

$$\begin{aligned} I_1 &= \text{tr} C = \lambda_1^2 + \lambda_2^2 + \lambda_3^2, \\ I_2 &= \text{tr}(\text{cof} C) = \frac{1}{2}\left(I_1^2 - \text{tr}(C^2)\right) = \lambda_1^2 \lambda_2^2 + \lambda_2^2 \lambda_3^2 + \lambda_1^2 \lambda_3^2, \\ I_3 &= \det C = \lambda_1^2 \lambda_2^2 \lambda_3^2, \end{aligned} \qquad (1)$$

where $\text{cof} C = \det(C) C^{-1}$ denotes the cofactor of right Cauchy Green deformation tensor $C$; $\text{tr}(\cdot)$ and $\det(\cdot)$ are trace and determinant operators, respectively. In the undeformed state, both the deformation gradient and the Cauchy-Green deformation tensor are identical to the unit tensor: $F = I$, $C = I$, and the Jacobian equals one, $J = 1$.

Furthermore, we introduce two types of stresses: the symmetric Cauchy stress $\boldsymbol{\sigma}$, denoting the force per deformed area along the outward normal direction $\boldsymbol{n}_s$, and the asymmetric first Piola-Kirchhoff stress $\boldsymbol{P}$, defined as the force per undeformed area along the outward normal direction $\boldsymbol{N}_s$. The transpose of the latter is also known as nominal stress, which is commonly employed as the stress measure in experiments. The relation between these two stresses is characterized by Piola transformation:

$$\boldsymbol{P} = J\boldsymbol{\sigma}\boldsymbol{F}^{-T} \quad \text{or} \quad \boldsymbol{\sigma} = J^{-1}\boldsymbol{P}\boldsymbol{F}^{T}. \tag{2}$$

In general, the second-order stress tensor $\boldsymbol{P}$ is not symmetric and has nine independent components. Left-multiplying $\boldsymbol{P}$ with $\boldsymbol{F}^{-1}$ gives its symmetric counterpart: the second Piola-Kirchhoff stress $\boldsymbol{S} = \boldsymbol{F}^{-1}\boldsymbol{P} = J\boldsymbol{F}^{-1}\boldsymbol{\sigma}\boldsymbol{F}^{-T}$.

The constitutive relationship establishes the connection between strain and stress in a material, reflecting the material response under external stimuli like applied forces or temperature variation. This relationship is a fundamental aspect of the material behavior and is commonly expressed in mathematical or tensorial form, as exemplified by the first Piola-Kirchhoff stress and the deformation gradient, $\boldsymbol{P} = \boldsymbol{P}(\boldsymbol{F})$. For hyperelastic materials, constitutive relations can be reformulated by positing the existence of a Helmholtz free-energy function, $\Psi(\boldsymbol{F})$.

## 2.2. Strain energy density function

The strain energy density function, i.e., the Helmholtz free energy under isothermal conditions ($\Psi$), provides an implicit mathematical combination of the strain and stress tensors, $\dot{\Psi} = \int_V \boldsymbol{A} : \dot{\boldsymbol{B}} \, \mathrm{d}V$, where $\boldsymbol{A}$ and $\boldsymbol{B}$ are stress and strain measures, respectively. The integrand $\boldsymbol{A} : \dot{\boldsymbol{B}}$ is identified as the stress power. In accordance with this form, we say that the stress field $\boldsymbol{A}$ is work conjugate to the strain field $\boldsymbol{B}$. This conjugacy implies a complementary relationship between stress and strain, providing an additional constraint for the selection of stress and strain pairs. A widely used conjugate pair is the first Piola-Kirchhoff stress $\boldsymbol{P}$ and the deformation gradient $\boldsymbol{F}$, resulting in the strain energy density function expressed as $\dot{\Psi} = \int_V \boldsymbol{P} : \dot{\boldsymbol{F}} \, \mathrm{d}V$. This implicit relationship highlights its effectiveness in deducing stress given information about $\Psi$ and $\boldsymbol{F}$. However, the validity of this application is reliant on two crucial considerations: the existence of Helmholtz free energy and adherence to the principles of physical admissibility. These aspects will be elaborated in the following subsections.

### 2.2.1. Thermodynamic consistency

Within the isothermal regime, fulfilling the second law of thermodynamics necessitates non-negative internal dissipation ($\mathcal{D}_{int}$), as expressed in Clausius-Planck inequality [45]:

$$\mathcal{D}_{int} = \boldsymbol{P}:\dot{\boldsymbol{F}} - \dot{\Psi}(\boldsymbol{F}) = \boldsymbol{P}:\dot{\boldsymbol{F}} - \frac{\partial \Psi}{\partial \boldsymbol{F}}:\dot{\boldsymbol{F}} \geq 0. \tag{3}$$

For hyperelastic or Green-elastic materials, the dissipation consistently remains zero, $\mathcal{D}_{int} \equiv 0$. Therefore, for an arbitrary choice of the tensor variable $\dot{\boldsymbol{F}}$, the first Piola-Kirchhoff stress can be expressed as:

$$\boldsymbol{P} = \frac{\partial \Psi(\boldsymbol{F})}{\partial \boldsymbol{F}}. \tag{4}$$

### 2.2.2. Material objectivity and frame indifference

The constitutive law must exhibit independence from observers or frames; thus, material objectivity must be satisfied by any strain energy function. This requirement necessitates the equivalence of $\Psi(\boldsymbol{F}) = \Psi(\boldsymbol{Q}\boldsymbol{F})$ for any arbitrary deformation gradient tensor $\boldsymbol{F}$ and proper orthogonal tensor $\boldsymbol{Q}$. Therefore, for the particular choice $\boldsymbol{Q} = \boldsymbol{R}^T$, where $\boldsymbol{R}$ represents the rotational part in the polar decomposition of $\boldsymbol{F}$ with $\boldsymbol{F} = \boldsymbol{R}\boldsymbol{U}$, where $\boldsymbol{U}$ is the stretch tensor, it is imperative that:

$$\Psi(\boldsymbol{F}) = \Psi(\boldsymbol{Q}\boldsymbol{F}) = \Psi(\boldsymbol{R}^T\boldsymbol{F}) = \Psi(\boldsymbol{U}). \tag{5}$$

Objectivity requires that the strain energy function $\Psi$ solely depends on stretching, and not on rotation or translation. Utilizing the relationship $\boldsymbol{U} = \sqrt{\boldsymbol{C}}$, the strain energy function $\Psi$ can be further articulated as a function of the six-component symmetric right Cauchy-Green tensor $\boldsymbol{C}$, namely $\Psi(\boldsymbol{F}) = \Psi(\boldsymbol{C})$. Accordingly, the first Piola-Kirchhoff stress $\boldsymbol{P}$ can be further detailed as:

$$\boldsymbol{P} = \frac{\partial \Psi(\boldsymbol{C})}{\partial \boldsymbol{F}} = \frac{\partial \Psi(\boldsymbol{C})}{\partial \boldsymbol{C}}:\frac{\partial \boldsymbol{C}}{\partial \boldsymbol{F}} = \frac{\partial \Psi(\boldsymbol{C})}{\partial \boldsymbol{C}}:\frac{\partial(\boldsymbol{F}^T\boldsymbol{F})}{\partial \boldsymbol{F}} = 2\boldsymbol{F}\cdot\frac{\partial \Psi(\boldsymbol{C})}{\partial \boldsymbol{C}}. \tag{6}$$

### 2.2.3. Material isotropy

For an isotropic material, the material response remains unchanged under the transformation of the reference configuration. Consequently, the equivalence $\Psi(\boldsymbol{C}) = \Psi(\boldsymbol{Q}\boldsymbol{C}\boldsymbol{Q}^T)$ holds for all orthogonal tensors $\boldsymbol{Q}$ and symmetric positive-definite tensors $\boldsymbol{C}$. Following Sylvester's law of inertia in algebra, the isotropic strain energy function $\Psi$ depends on $C$ only through its principal scalar invariants defined in Equation (1). This implies the existence of a function $\widehat{\Psi}$ such that $\Psi(C) = \widehat{\Psi}(I_1(C), I_2(C), I_3(C))$, and the first Piola-Kirchhoff stress $P$ can be further represented as:

$$P = \frac{\partial \widehat{\Psi}(I_1, I_2, I_3)}{\partial F} = 2\left[\frac{\partial \widehat{\Psi}}{\partial I_1} + I_1 \frac{\partial \widehat{\Psi}}{\partial I_2}\right] F - 2\frac{\partial \widehat{\Psi}}{\partial I_2} F \cdot C + 2I_3 \frac{\partial \widehat{\Psi}}{\partial I_3} F^{-T}. \tag{7}$$

Analogously, we can also express $\Psi$ in terms of the three principal stretches of the deformation gradient $F$. As such, the first Piola-Kirchhoff stress $P$ takes the following form:

$$P = \sum_{i=1}^{3} \frac{\partial \Psi}{\partial \lambda_i} n_i \otimes N_i. \tag{8}$$

### 2.2.4. Incompressibility

For the case of perfect incompressibility, the Jacobian remains constant and equal to one, $I_3 = J^2 = 1$. The strain energy function proposed for such constrained materials is postulated to adopt the form $\Psi = \Psi(F) - p(J-1)$, where the scalar $p$ serves as an indeterminate Lagrange multiplier, also identified as the hydrostatic pressure. Upon differentiating this function with respect to the deformation gradient $F$ and employing the formula $\partial J / \partial F = J F^{-T}$, we obtain the following expression for the first Piola-Kirchhoff stress $P$:

$$P = \frac{\partial \Psi(F)}{\partial F} - pF^{-T} = 2\left[\frac{\partial \Psi}{\partial I_1} + I_1 \frac{\partial \Psi}{\partial I_2}\right] F - 2\frac{\partial \Psi}{\partial I_2} F \cdot C - pF^{-T}. \tag{9}$$

In the spectral representation, the resultant first Piola-Kirchhoff stress $P$ is given as:

$$P = \sum_{i=1}^{3} \frac{\partial \Psi}{\partial \lambda_i} n_i \otimes N_i - pF^{-T}. \tag{10}$$

### 2.2.5. Polyconvexity and coercivity condition

For a boundary value problem in nonlinear elasticity, the existence and uniqueness of the solution are intrinsically related to the convexity of the strain energy function $\Psi$. Convexity plays a pivotal role as it implies ellipticity, thereby assuring material stability in a constitutive model by prescribing convex shapes [19]. Moreover, convexity ensures that the energy function exclusively attains its global minimum at thermodynamic equilibrium within the reference configuration. However, general convexity is often considered too restrictive from physical perspectives and proves challenging to impose in practical mathematical applications [16]. For example, enforcing convexity with respect to the deformation gradient $\boldsymbol{F}$ leads to the following inequality:

$$\Psi(\boldsymbol{F}) - \Psi(\boldsymbol{F}_0) \geq \left.\frac{\partial \Psi}{\partial \boldsymbol{F}}\right|_{\boldsymbol{F}_0} : (\boldsymbol{F} - \boldsymbol{F}_0), \tag{11}$$

for all admissible deformation gradients $\boldsymbol{F}$ and their neighboring parts $\boldsymbol{F}_0$. Instead, a less restrictive requirement is the polyconvexity of the strain energy function [28, 46]. Polyconvexity of $\Psi(\boldsymbol{F})$ demands sufficient convexity concerning the extended domain formed by deformation gradient $\boldsymbol{F}$, its cofactor $\text{cof}\boldsymbol{F}$, and determinant $\det\boldsymbol{F}$ [47]. Therefore, there exists a representative strain energy function $\widehat{\Psi}(\boldsymbol{F})$ such that

$$\Psi(\boldsymbol{F}) = \widehat{\Psi}(\boldsymbol{F}, \text{cof}\boldsymbol{F}, \det\boldsymbol{F}). \tag{12}$$

Constructing a general function that precisely fulfills this requirement can be challenging, and a more flexible and pragmatic approach is to find subsets through the additive decomposition [48],

$$\widehat{\Psi}(\boldsymbol{F}, \text{cof}\boldsymbol{F}, \det\boldsymbol{F}) = \widehat{\Psi}_{\text{F}}(\boldsymbol{F}) + \widehat{\Psi}_{\text{cof}}(\text{cof}\boldsymbol{F}) + \widehat{\Psi}_{\text{det}}(\det\boldsymbol{F}). \tag{13}$$

where $\widehat{\Psi}_{\text{F}}$, $\widehat{\Psi}_{\text{cof}}$, and $\widehat{\Psi}_{\text{det}}$ are convex function with respect to $\boldsymbol{F}$, $\text{cof}\boldsymbol{F}$, and $\det\boldsymbol{F}$, respectively. Above considerations of polyconvexity all pertain to the deformation gradient $\boldsymbol{F}$, however, the convexity with respect to $\boldsymbol{F}$ encounters incompatible with the principal of objectivity and is not suitable for finite elasticity. To address this limitation, we reformulate the polyconvexity condition by involving the invariants of the right Cauchy Green deformation tensor $\boldsymbol{C}$. Moreover, it is noteworthy that non-decreasing substitutions of invariants, $I_1 = \text{tr}\boldsymbol{C}$, $I_2 = \text{tr}(\text{cof}\boldsymbol{C})$, $I_3 = \det\boldsymbol{C}$, preserve convexity [44]. Therefore, the strain energy function can be further simplified as the summation of invariant-based functions,

$$\widehat{\Psi}_{\text{F}}(\boldsymbol{F}) + \widehat{\Psi}_{\text{cof}}(\text{cof}\boldsymbol{F}) + \widehat{\Psi}_{\text{det}}(\det\boldsymbol{F}) = \widehat{\Psi}_{I_1}(I_1) + \widehat{\Psi}_{I_2}(I_2) + \widehat{\Psi}_{I_3}(I_3). \tag{14}$$

where $\widehat{\Psi}_{I_1}$, $\widehat{\Psi}_{I_2}$, $\widehat{\Psi}_{I_3}$ are convex functions with respect to the three invariants. Accounting for incompressibility, the contribution of $I_3$ can be neglected due to its constant value. Consequently, polyconvexity enforces the forms of the strain energy function as $\Psi(\boldsymbol{F}) = \widehat{\Psi}_{I_1}(I_1) + \widehat{\Psi}_{I_2}(I_2)$.

An alternative and less restrictive convexity condition is rank-one convexity, also known as the ellipticity condition. The strain energy function remains elliptic when the Legendre-Hadamard condition holds [16],

$$(\boldsymbol{M}\otimes\boldsymbol{m}):\frac{\partial^2\Psi}{\partial\boldsymbol{F}^2}:(\boldsymbol{M}\otimes\boldsymbol{m}) \geq 0, \tag{15}$$

where $\boldsymbol{M}$ and $\boldsymbol{m}$ denote arbitrary vectors in the reference or material and current or spatial configurations. The ellipticity condition implies positive-semi-definiteness of the tangent tensor $\partial^2\Psi/\partial\boldsymbol{F}^2$ [11, 49]. Particularly, when the vectors are coaxial with the principal direction of the deformation gradient, i.e., $\boldsymbol{N}$ and $\boldsymbol{n}$, this condition simplifies to a positivity check for the following Hessian matrix,

$$\boldsymbol{H} = \begin{bmatrix} \frac{\partial^2\Psi}{\partial\lambda_1^2} & \frac{\partial^2\Psi}{\partial\lambda_1\partial\lambda_2} & \frac{\partial^2\Psi}{\partial\lambda_1\partial\lambda_3} \\ \frac{\partial^2\Psi}{\partial\lambda_2\partial\lambda_1} & \frac{\partial^2\Psi}{\partial\lambda_2^2} & \frac{\partial^2\Psi}{\partial\lambda_2\partial\lambda_3} \\ \frac{\partial^2\Psi}{\partial\lambda_3\partial\lambda_1} & \frac{\partial^2\Psi}{\partial\lambda_3\partial\lambda_2} & \frac{\partial^2\Psi}{\partial\lambda_3^2} \end{bmatrix}, \tag{16}$$

here, $\lambda_1$, $\lambda_2$, $\lambda_3$ are the three eigenvalues of the deformation gradient $\boldsymbol{F}$, corresponding to the principal stretches along three principal directions. Therefore, if the three eigenvalues are positive and the determinant of $\boldsymbol{H}$ is positive, the ellipticity condition is fulfilled for the strain energy function $\Psi(\boldsymbol{F})$. Analogously, the additive decomposition can be applied to refine the available function subsets:

$$\Psi(\lambda_1, \lambda_2, \lambda_3) = \Psi_1(\lambda_1) + \Psi_2(\lambda_2) + \Psi_3(\lambda_3). \tag{17}$$

This decomposition ensures that all off-diagonal components of the Hessian matrix are zero, and its three eigenvalues directly correspond to the diagonal components: $\partial^2\Psi/\partial\lambda_1^2$, $\partial^2\Psi/\partial\lambda_2^2$, $\partial^2\Psi/\partial\lambda_3^2$. Notably, the polyconvexity condition is slightly more restrictive than the rank-one convexity condition. However, both conditions are sufficient conditions for the existence of minimizers, indicating the potential existence of multiple local minimizers under each condition. To address this issue, a coercivity condition can be introduced. The coercivity condition, also known as the growth condition, demands that the stresses should grow unboundedly with infinite strains [44, 50]. For simplicity, the strain energy density $\Psi$ is considered infinite for infinite compression, $J \to 0$, and infinite expansion, $J \to \infty$.

In the current study, the polyconvexity condition, along with the coercivity condition, is employed to determine the existence and uniqueness of the invariant-based strain energy function $\Psi(I_1, I_2)$, while the rank-one convexity condition, along with the coercivity condition, is utilized to verify the existence of global minimizers for the principal stretch-based or strain-based strain energy function $\Psi(\lambda_1, \lambda_2, \lambda_3)$.

### 2.2.6. Non-negativeness and normalization condition

Intuitively, the strain energy function $\Psi$ should be non-negative for all deformation states, and converge to zero in the reference configuration, when $\boldsymbol{F} = \boldsymbol{I}$:

$$\begin{aligned} \Psi(\boldsymbol{F}) \geq 0 \quad &\forall \boldsymbol{F}, \\ \Psi(\boldsymbol{I}) = 0, \boldsymbol{P}(\boldsymbol{I}) &= 0. \end{aligned} \tag{18}$$

### 2.3. Symbolic regression

Symbolic regression stands out as a distinctive form of regression, wherein a mathematic expression is autonomously identified to best fit the provided dataset. Unlike conventional regression or data-driven methods that require predefined model structures, such as artificial neural networks (ANN), symbolic regression is capable of generating analytical expressions purely from data without the specific need of prior knowledge, thereby significantly enhancing the interpretability, generalizability, and flexibility of the model discovery process [39].

The algorithm for symbolic regression unfolds in an evolutional manner, known as genetic programming (GP), which draws inspirations from the Darwinian principles of natural selection. Within GP, functional expressions are efficiently represented using a binary-tree structure, comprising nodes and branches, as illustrated in Figure 1a. A complete tree structure involves variables, mathematic operators, either unary or binary, and constants.

In the initial stages, the algorithm randomly generates a population of symbolic tree expressions based on user-defined variables and operators, serving as candidate functionals. For each candidate expression, the fitness is evaluated through the calculation of the mean square error (MSE) between predicted outputs and target values. Expressions with higher fitness values are more likely to be selected as baselines for subsequent optimization, where the expression trees are modified primarily through two genetic operations: mutation and crossover. The mutation operation entails randomly altering some nodes in an expression tree, introducing genetic diversity in the population. An example is shown in Figure 1b, where a new offspring is generated by replacing the unary operator "log" with "sin". On the other hand, the crossover operation permits the algorithms to create new offspring by combining building blocks from different parent individuals, as demonstrated in Figure 1c. These newly generated expressions become candidates for the next iteration. The iterative process of evaluation, selection, mutation, and crossover constitutes one evolution of the regression analysis. This cycle is repeated until the optimal expression is obtained or the maximum number of generations is reached [51, 52].

To enhance interpretability and mitigate potential overfitting, achieving a balance between model accuracy and complexity is crucial. However, there is still a lack of consensus on the precise definition of these two terms. Herein, we adopt a proposed measure [53] that defines complexity as the number of nodes in an expression tree. The loss of expression $\mathcal{L}(Expr)$ is then evaluated as a combination of the predictive loss $\mathcal{L}_{pred}(Expr)$ and the complexity measure $C(Expr)$,

$$\mathcal{L}(Expr) = \mathcal{L}_{pred}(Expr) \cdot \exp(\text{frec}[C(Expr)]), \tag{19}$$

where $\text{frec}[C(Expr)]$ represents a combined measure of the frequency and recency of expressions occurring at the current complexity in the population [53]. The optimal expression is determined based on a score metric, as the negated derivative of the log-loss with respect to the complexity, $-\text{d}[\log(\text{MAE})]/\text{d}C$.

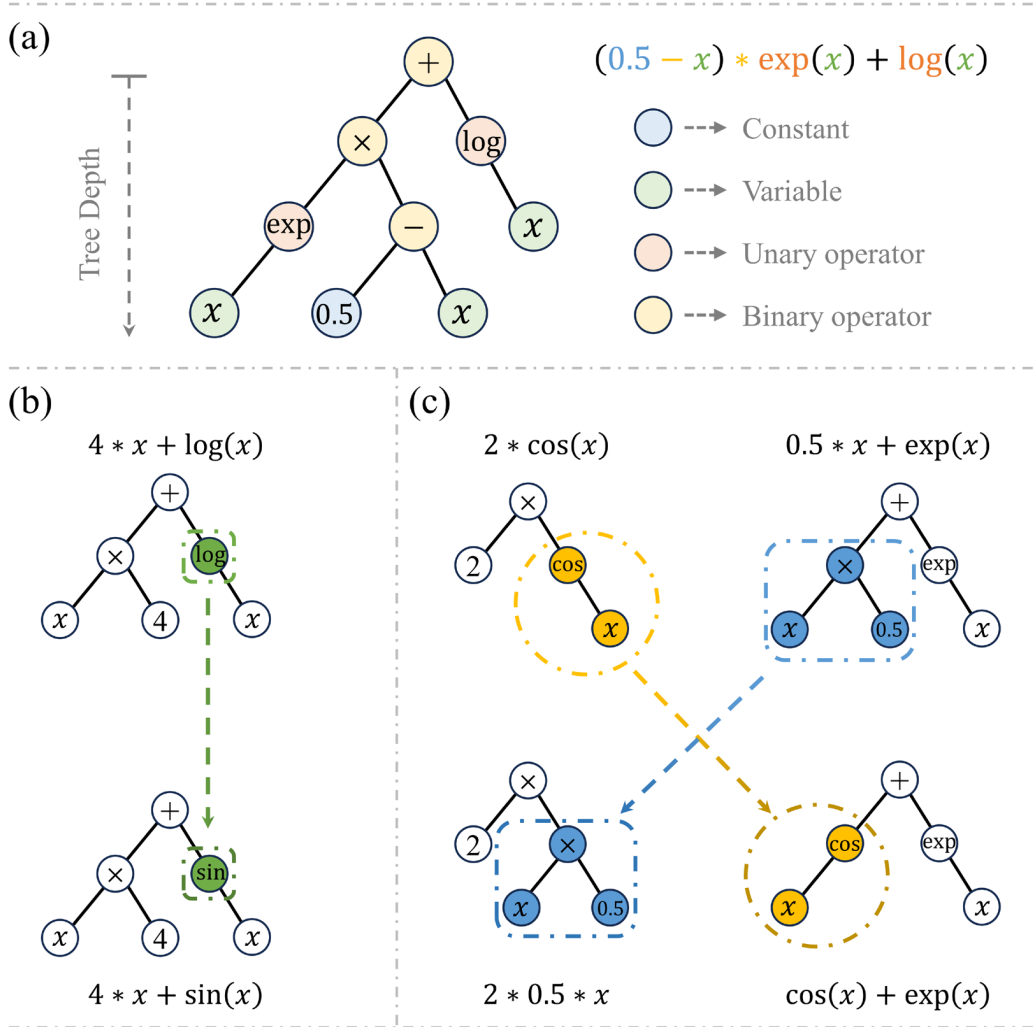

**Figure 1. Structure and operations of expression tree**. (a). Representation of expression tree for an algebraic expression $((0.5 - x) * \exp(x) + \log(x))$; (b). An example of the mutation operation; (c). An example of the crossover operation.

In this manuscript, we employ the symbolic regression algorithms introduced above to predict the optimal strain energy function for human brain tissue, leveraging multi-mode experimental data in uniaxial tension, uniaxial compression, and simple shear. To explore the capabilities of symbolic regression, we incorporate three distinct sets of variables as model inputs: the invariants $(I_1, I_2)$, principal stretches $(\lambda_1, \lambda_2, \lambda_3)$, and principal strains $(\epsilon_1, \epsilon_2, \epsilon_3)$. The corresponding strain energy functions we seek to discover are $\Psi(I_1, I_2)$, $\Psi(\lambda_1, \lambda_2, \lambda_3)$, $\Psi(\epsilon_1, \epsilon_2, \epsilon_3)$, respectively. For clarity, we term the related algorithms as "*Invariant-based Symbolic Regression*", "*Stretch-based*

*Symbolic Regression*", and "*Strain-based Symbolic Regression*". In each algorithm, we meticulously craft the model structure and objective functions to ensure the physical admissibility of the strain energy functions by adhering to the physical constraints outlined in Section 2.2.

### 2.3.1. Invariant-based Symbolic Regression

*Invariant-based Symbolic Regression* takes the invariants $(I_1, I_2, I_3)$ of the right Cauchy-Green tensor $\boldsymbol{C}$ as input and the strain energy function $\Psi$ as output. The data used to train these symbolic regression models for the human brain consist of three loading modes, uniaxial tension, uniaxial compression, and simple shear [3]. The experimental data include the deformation gradient $\boldsymbol{F}$ and the first Piola-Kirchhoff stresses $\boldsymbol{P}$. Below, we briefly introduce the mathematical representations of these quantities based on the experimental measures for each loading mode.

**Uniaxial tension and compression.** In the scenario of unconfined uniaxial tension and compression tests, the specimen undergoes uniform deformation along the stretching direction. Under the assumptions of material isotropy and perfect incompressibility, the deformation gradient $\boldsymbol{F}$ and right Cauchy Green deformation tensor $\boldsymbol{C}$ take the following matrix forms, described in terms of the applied stretch $\lambda$,

$$[F] = \begin{bmatrix} \lambda & 0 & 0 \\ 0 & 1/\sqrt{\lambda} & 0 \\ 0 & 0 & 1/\sqrt{\lambda} \end{bmatrix} \qquad [C] = \begin{bmatrix} \lambda^2 & 0 & 0 \\ 0 & 1/\lambda & 0 \\ 0 & 0 & 1/\lambda \end{bmatrix}, \tag{20}$$

with three invariants

$$I_1 = \lambda^2 + \frac{2}{\lambda} \qquad I_2 = 2\lambda + \frac{1}{\lambda^2} \qquad I_3 = 1. \tag{21}$$

Using Equation (9), the nominal uniaxial stress can be calculated as:

$$P_{11} = 2\left(\frac{\partial \Psi}{\partial I_1} + \frac{1}{\lambda}\frac{\partial \Psi}{\partial I_2}\right)\left(\lambda - \frac{1}{\lambda^2}\right) \qquad P_{22} = P_{33} = 0. \tag{22}$$

Notably, the hydrostatic pressure $p$ is determined from the free stress assumption in the transverse plane, occurring in the unconfined test, namely, $P_{22} = 0$.

**Simple shear.** During the simple shear test, a specified amount of shear ($\gamma$) is applied in the x-y plane of an isotropic and perfectly incompressible specimen. The resultant deformation gradient $\boldsymbol{F}$ and right Cauchy Green deformation tensor $\boldsymbol{C}$ take the following matrix form:

$$[F] = \begin{bmatrix} 1 & \gamma & 0 \\ 0 & 1 & 0 \\ 0 & 0 & 1 \end{bmatrix} \quad [C] = \begin{bmatrix} 1 & \gamma & 0 \\ \gamma & 1+\gamma^2 & 0 \\ 0 & 0 & 1 \end{bmatrix}, \tag{23}$$

with three invariants,

$$I_1 = 3 + \gamma^2 \quad I_2 = 3 + \gamma^2 \quad I_3 = 1. \tag{24}$$

The hydrostatic pressure $p$ has no impact on the off-diagonal nominal shear stress $P_{12}$, leading to the following explicit form:

$$P_{12} = 2\left(\frac{\partial \Psi}{\partial I_1} + \frac{\partial \Psi}{\partial I_2}\right)\gamma. \tag{25}$$

During regression, all three loading modes are simultaneously employed to train the strain energy function model. This combined training approach enhances the model's ability to characterize material responses under complex loading situations, thereby improving the generalizability of the predicted model [33, 54]. In this manuscript, we employ the normalized mean square error to evaluate the loss between the predicted stress $P^*$ and the experimental stresses $P$,

$$L_{pred} = \frac{1}{N_{ut}} \sum_{i=1}^{N_{ut}} \left\| \frac{P_{ut,i} - P^*_{ut,i}}{P^{max}_{ut}} \right\|^2 + \frac{1}{N_{uc}} \sum_{i=1}^{N_{uc}} \left\| \frac{P_{uc,i} - P^*_{uc,i}}{P^{max}_{uc}} \right\|^2 + \frac{1}{N_{ss}} \sum_{i=1}^{N_{ss}} \left\| \frac{P_{ss,i} - P^*_{ss,i}}{P^{max}_{ss}} \right\|^2. \tag{26}$$

Here, each loss term is normalized by the maximum experimental stress ($P^{max}_{ut}$, $P^{max}_{uc}$, $P^{max}_{ss}$) to mitigate the impact introduced by the choice of stress measure [55, 56]. $P_{ut} = P_{11}$ for tension ($\lambda > 1$), $P_{uc} = P_{11}$ for tension ($\lambda < 1$), and $P_{ss} = P_{12}$ for simple shear ($\gamma > 0$).

The data fed into *Invariant-based Symbolic Regression* involve the invariants ($I_1, I_2$) and stresses ($P_{ut}$, $P_{uc}$, $P_{ss}$). Performing symbolic regression on these data directly yields a relation between stresses and invariants, such as $P_{ut}(I_1, I_2)$. However, thermodynamic consistency dictates an indirect relation, $\mathbf{P} = \partial \Psi / \partial \mathbf{F}$, implying that the target output of symbolic regression should be the strain energy density $\Psi(I_1, I_2)$. To address this, we customize the objective function to enable auto-differentiation inside the loss function, as shown in Algorithm 1. At each step, the derivatives of $\Psi$ with respect to $I_1$ and $I_2$ are calculated and stored as $\partial \Psi / \partial I_1$ and $\partial \Psi / \partial I_2$. These two derivatives are essential for determining the first Piola-Kirchhoff stresses, as indicated in Equations (22) and (25). Furthermore, to ensure a stress-free reference configuration, both invariants are shifted by 3 at the initial stage. Regarding the polyconvexity condition, we enforce all the constants to be non-negative [27, 44].

---

**Algorithm 1** Framework of the Customized Loss Function for *Invariant-based Symbolic Regression*

---

**Input:** First and second invariant calculated from experimental stretchs of uniaxial tension, uniaxial compression, and simple shear, $I_1$,$I_2$; First Piola-Kirchhoff stress from tension $P_{ut}$, compression $P_{uc}$, and shear $P_{ss}$

**Output:** Normolized mean square error, $L_{pred}$;

1: Shift $I_1$ and $I_2$ with 3 to ensure stress-free state at initial configuration;
2: Concatenate the inputs of three loading modes data along the raw direction;
3: Constrain all the constants to be non-negative;
4: Calculate the derivatives of strain energy density $w.r.t$ invariants, $\frac{\partial \Psi}{\partial I_1}, \frac{\partial \Psi}{\partial I_2}$;
5: Calculate the stretches or shear for each loading mode, $\lambda_{ut}, \lambda_{uc}, \gamma_{ss}$;
6: Determine the predicted first Piola-Kirchhoff stress: $P^*_{ut} = 2(\lambda_{ut} - \frac{1}{\lambda^2_{ut}})(\frac{\partial \Psi}{\partial I_1} + \frac{1}{\lambda_{ut}}\frac{\partial \Psi}{\partial I_2})$, $P^*_{uc} = 2(\lambda_{uc} - \frac{1}{\lambda^2_{uc}})(\frac{\partial \Psi}{\partial I_1} + \frac{1}{\lambda_{uc}}\frac{\partial \Psi}{\partial I_2})$, $P^*_{ss} = 2(\frac{\partial \Psi}{\partial I_1} + \frac{\partial \Psi}{\partial I_2})\gamma_{ss}$;
7: Evaluate the loss: $L_{pred} = \frac{1}{N_{ut}}\sum_{i=1}^{N_{ut}}\|\frac{P_{ut,i}-P^*_{ut,i}}{P^{max}_{ut}}\|^2 + \frac{1}{N_{uc}}\sum_{i=1}^{N_{uc}}\|\frac{P_{uc,i}-P^*_{uc,i}}{P^{max}_{uc}}\|^2 + \frac{1}{N_{ss}}\sum_{i=1}^{N_{ss}}\|\frac{P_{ss,i}-P^*_{ss,i}}{P^{max}_{ss}}\|^2$;
8: **return** $L_{pred}$.

---

In symbolic regression, the expression tree theoretically can take an arbitrary functional shape. However, in consideration of the computational costs, we constrain the evolving expressions to be within the domain constructed by polynomial, exponential, and logarithmic functions. These forms are commonly utilized in the classical hyperelastic models, such as Mooney Rivlin model [57], Gent model [58], and Holzapfel model [59]. The detailed model setups are summarized in Table 1. Note that, the complexities and constraints can be flexibly tuned for specific problems. For example, we can adjust the complexity of the multiplication operator ("*") to a much larger value than that of addition operator ("+") if we need to restrict the use of "*" in functional evolution.

### 2.3.2. Stretch-based Symbolic Regression

*Stretch-based Symbolic Regression* takes the principal stretches ($\lambda_1, \lambda_2, \lambda_3$) of the deformation gradient $\boldsymbol{F}$ as input and the strain energy function $\Psi(\lambda_1, \lambda_2, \lambda_3)$ as output. For this particular regression, we confine the functional format strictly to follow the generalized Ogden model [60],

$$\Psi(\lambda_1, \lambda_2, \lambda_3) = \sum_{k=1}^{n}\frac{\mu_k}{\alpha_k^2}[\lambda_1^{\alpha_k} + \lambda_2^{\alpha_k} + \lambda_3^{\alpha_k} - 3], \tag{27}$$

where $\mu_k$ represents the shear stiffness, $\alpha_k$ the nonlinearity parameter. Prior to application, the first step is to convert the experimental measure of the deformation gradient into its spectral representation, $[\mathbf{F}] = \text{diag}(\lambda_1, \lambda_2, \lambda_3)$.

**Uniaxial tension and compression.** For the special case of unconfined uniaxial tension and compression, the principal stretches can be expressed in terms of the prescribe stretch $\lambda$,

$$\lambda_1 = \lambda \quad \lambda_2 = 1/\sqrt{\lambda} \quad \lambda_3 = 1/\sqrt{\lambda}. \tag{28}$$

Using Equation (10) and free stress assumption, the nominal uniaxial stress can be calculated as:

$$P_{11} = \frac{\partial \Psi}{\partial \lambda_1} - \frac{1}{\lambda_1 \sqrt{\lambda_1}} \frac{\partial \Psi}{\partial \lambda_2}. \tag{29}$$

**Simple shear.** For the simple shear test prescribed in the x-y plane, the principal stretches are determined in terms of the shear strain $\gamma$,

$$\lambda_1 = \frac{\gamma + \sqrt{4 + \gamma^2}}{2} \quad \lambda_2 = \frac{-\gamma + \sqrt{4 + \gamma^2}}{2} \quad \lambda_3 = 1. \tag{30}$$

Using Equation (10), the shear stress has the following succinct expression:

$$P_{12} = \frac{\lambda_1^2}{\lambda_1^2 + 1} \frac{\partial \Psi}{\partial \lambda_1} - \frac{\lambda_2^2}{\lambda_2^2 + 1} \frac{\partial \Psi}{\partial \lambda_2}. \tag{31}$$

For a detailed mathematic derivation of Equation (31), please refer to the *Appendix B* in our recently published paper [56].

---

**Algorithm 2** Framework of the Customized Loss Function for *Stretch-based Symbolic Regression*

---

**Input:** Principal stretchs of uniaxial tension $\lambda_{k,ut}$, uniaxial compression $\lambda_{k,uc}$, and simple shear $\lambda_{k,ss}$, with $k$ in $1, 2, 3$; First Piola-Kirchhoff stress from tension $P_{ut}$, compression $P_{uc}$, and shear $P_{ss}$;

**Output:** Normolized mean square error, $L_{pred}$;

1: Concatenate all the principal stretches and stresses into one column, respectively;
2: Calculate the derivatives of strain energy density $w.r.t$ principal stretches, $\frac{\partial \Psi}{\partial \lambda_1}, \frac{\partial \Psi}{\partial \lambda_2}, \frac{\partial \Psi}{\partial \lambda_3}$;
3: Determine the predicted first Piola-Kirchhoff stress: $P_{ut}^* = \frac{\partial \Psi}{\partial \lambda_1} - \frac{\lambda_{2,ut}}{\lambda_{1,ut}} \frac{\partial \Psi}{\partial \lambda_2})$, $P_{uc}^* = \frac{\partial \Psi}{\partial \lambda_1} - \frac{\lambda_{2,uc}}{\lambda_{1,uc}} \frac{\partial \Psi}{\partial \lambda_2})$, $P_{ss}^* = \frac{\lambda_{1,ss}^2}{\lambda_{1,ss}^2+1} \frac{\partial \Psi}{\partial \lambda_1} - \frac{\lambda_{2,ss}^2}{\lambda_{2,ss}^2+1} \frac{\partial \Psi}{\partial \lambda_2})$;
4: Evaluate the loss: $L_{pred} = \frac{1}{N_{ut}} \sum_{i=1}^{N_{ut}} \|\frac{P_{ut,i} - P_{ut,i}^*}{P_{ut}^{max}}\|^2 + \frac{1}{N_{uc}} \sum_{i=1}^{N_{uc}} \|\frac{P_{uc,i} - P_{uc,i}^*}{P_{uc}^{max}}\|^2 + \frac{1}{N_{ss}} \sum_{i=1}^{N_{ss}} \|\frac{P_{ss,i} - P_{ss,i}^*}{P_{ss}^{max}}\|^2$;
5: **return** $L_{pred}$.

Analogously, for *Stretch-based Symbolic Regression*, we employ the normalized mean square error to evaluate the loss between the predicted stress $P^*$ and experimental stresses $P$, with the expression identical to Equation (26). Again, we customize the objective function to ensure the thermodynamic consistency, with its framework depicted in Algorithm 2.

At each step, the derivatives of $\Psi$ with respect to $\lambda_1$, $\lambda_2$, and $\lambda_3$ are calculated and stored in $\partial\Psi/\partial\lambda_1$, $\partial\Psi/\partial\lambda_2$, and $\partial\Psi/\partial\lambda_3$, respectively. These derivatives further contribute to determining the first Piola-Kirchhoff stresses $P_{ut}^*$, $P_{uc}^*$, and $P_{ss}^*$, as indicated in Equations (29) and (31). In the algorithm, the first step is to concatenate all the principal stretches and stresses into one column, respectively. This tricky recombination is crucial for code execution because the expression tree is constrained to follow the format of Ogden model. Thus, differentiation operations of $\Psi$ with respect to each principal stretch share equal weight. This allows us to simplify the derivatives of $\Psi$ with respect to a single variable, e.g., $\left.\frac{\partial\Psi}{\partial\lambda}\right|_{\lambda=\lambda_1,\lambda_2,\lambda_3}$. Notably, the polyconvexity condition is not enforced as described for the *Invariant-based Symbolic Regression* because the nonlinearity parameter $\alpha_k$ is allowed to be negative. However, the coefficient $\mu_k/\alpha_k^2$ must be strictly positive to ensure the positivity of shear stiffness [61]. The polyconvexity condition will be post-checked as described in Section 2.2.5. Ultimately, the evolving expressions are restricted to polynomial functions. The detailed model setups are summarized in Table 1.

### 2.3.3. Strain-based Symbolic Regression

*Strain-based Symbolic Regression* takes the principal strains $(\epsilon_1, \epsilon_2, \epsilon_3)$ of the deformation gradient $F$ as input and the strain energy function $\Psi(\epsilon_1, \epsilon_2, \epsilon_3)$ as output. The principal strains are also referred to as Biot strains, representing the strain measure in the normal direction. The relation between Biot strain and principal stretch is described as:

$$\epsilon_i = \lambda_i - 1, \quad \text{with } i \text{ in } 1, 2, 3 \tag{32}$$

In contrast to *Stretch-based Symbolic Regression*, where an Ogden functional format is specified for the expression tree, *Strain-based Symbolic Regression* considers the strain energy function as a polynomial expansion of the Biot strain measure [56],

$$\Psi(\epsilon_1, \epsilon_2, \epsilon_3) = \sum_{k=1}^{n} \beta_k \left(\epsilon_1^k + \epsilon_2^k + \epsilon_3^k\right). \tag{33}$$

Using the relation defined in Equation (32), we can reformulate the mathematic representation of stresses in terms of the principal strains $(\epsilon_1, \epsilon_2, \epsilon_3)$

**Uniaxial tension and compression.** For the special case of unconfined uniaxial tension and compression, the principal strains can be expressed in terms of the prescribe stretch $\lambda$,

$$\epsilon_1 = \lambda - 1 \quad \epsilon_2 = 1/\sqrt{\lambda} - 1 \quad \epsilon_3 = 1/\sqrt{\lambda} - 1. \tag{34}$$

Reshaping the Equation (29) using the chain rule, the nominal uniaxial stress can be determined as:

$$P_{11} = \frac{\partial \Psi}{\partial \epsilon_1} - \frac{1}{(\epsilon_1 + 1)\sqrt{(\epsilon_1 + 1)}} \frac{\partial \Psi}{\partial \epsilon_2}. \tag{35}$$

**Simple shear.** For the simple shear test prescribed in the x-y plane, the principal strains are expressed in terms of the shear strain $\gamma$,

$$\epsilon_1 = \frac{\gamma + \sqrt{4 + \gamma^2}}{2} - 1 \quad \epsilon_2 = \frac{-\gamma + \sqrt{4 + \gamma^2}}{2} - 1 \quad \epsilon_3 = 0. \tag{36}$$

Analogously, reformulating Equation (31) yields the expression for shear stress:

$$P_{12} = \frac{(\epsilon_1 + 1)^2}{(\epsilon_1 + 1)^2 + 1} \frac{\partial \Psi}{\partial \epsilon_1} - \frac{(\epsilon_2 + 1)^2}{(\epsilon_2 + 1)^2 + 1} \frac{\partial \Psi}{\partial \epsilon_2}. \tag{37}$$

For *Strain-based Symbolic Regression*, we also use the normalized mean square error to assess the loss between the predicted stress $\boldsymbol{P}^*$ and experimental stresses $\boldsymbol{P}$, leveraging the expression defined in Equation (26). Additionally, we customize the objective function to ensure thermodynamic consistency, as outlined in Algorithm 3. In each iteration, the derivatives of $\Psi$ with respect to $\epsilon_1$, $\epsilon_2$, and $\epsilon_3$ are calculated and stored in $\partial \Psi/\partial \epsilon_1$, $\partial \Psi/\partial \epsilon_2$, and $\partial \Psi/\partial \epsilon_3$, respectively. Furthermore, concatenation is also performed to facilitate the differentiation operation. Herein, we impose no restriction on the sign of $\beta_k$, however, the positivity of shear modulus is strictly enforced by considering the consistency condition, namely, the isotropic hyperelastic model should be consistent with linear elasticity theory for small strains [45],

$$\mu = \frac{1}{2}\left[\frac{\partial^2 \Psi(0,0,0)}{\partial \epsilon_i^2} - \frac{\partial^2 \Psi(0,0,0)}{\partial \epsilon_i \partial \epsilon_j} + \frac{\partial \Psi(0,0,0)}{\partial \epsilon_i}\right] > 0. \tag{38}$$

Here, $\Psi(0,0,0)$ indicates the derivates of $\Psi(\epsilon_1, \epsilon_2, \epsilon_3)$ in the reference configuration where all principal strains have a constant value of 0. The polyconvexity condition will be post-checked as described in Section 2.2.5, and the evolving expressions are confined to polynomial functions. The model setups closely resemble those of the *Stretch-based Symbolic Regression*, as described in Table 1.

---

**Algorithm 3** Framework of the Customized Loss Function for *Strain-based Symbolic Regression*

**Input:** Principal strains of uniaxial tension $\epsilon_{k,ut}$, uniaxial compression $\epsilon_{k,uc}$, and simple shear $\epsilon_{k,ss}$, with $k$ in $1, 2, 3$; First Piola-Kirchhoff stress from tension $P_{ut}$, compression $P_{uc}$, and shear $P_{ss}$;
**Output:** Normolized mean square error, $L_{pred}$;
1: Concatenate all the principal strains and stresses into one column, respectively;
2: Calculate the derivatives of strain energy density $w.r.t$ principal strains, $\frac{\partial \Psi}{\partial \epsilon_1}, \frac{\partial \Psi}{\partial \epsilon_2}, \frac{\partial \Psi}{\partial \epsilon_3}$;
3: Derive the principal stretches based on given principal strains $\lambda_{k,ut} = \epsilon_{k,ut} + 1$, $\lambda_{k,uc} = \epsilon_{k,uc} + 1$, $\lambda_{k,ss} = \epsilon_{k,ss} + 1$, with $k$ in $1, 2, 3$;
4: Determine the predicted first Piola-Kirchhoff stress: $P_{ut}^* = \frac{\partial \Psi}{\partial \epsilon_1} - \frac{\lambda_{2,ut}}{\lambda_{1,ut}}\frac{\partial \Psi}{\partial \epsilon_2}$), $P_{uc}^* = \frac{\partial \Psi}{\partial \epsilon_1} - \frac{\lambda_{2,uc}}{\lambda_{1,uc}}\frac{\partial \Psi}{\partial \epsilon_2}$), $P_{ss}^* = \frac{\lambda_{1,ss}^2}{\lambda_{1,ss}^2+1}\frac{\partial \Psi}{\partial \epsilon_1} - \frac{\lambda_{2,ss}^2}{\lambda_{2,ss}^2+1}\frac{\partial \Psi}{\partial \epsilon_2}$);
5: Evaluate the loss: $L_{pred} = \frac{1}{N_{ut}}\sum_{i=1}^{N_{ut}}\|\frac{P_{ut,i}-P_{ut,i}^*}{P_{ut}^{max}}\|^2 + \frac{1}{N_{uc}}\sum_{i=1}^{N_{uc}}\|\frac{P_{uc,i}-P_{uc,i}^*}{P_{uc}^{max}}\|^2 + \frac{1}{N_{ss}}\sum_{i=1}^{N_{ss}}\|\frac{P_{ss,i}-P_{ss,i}^*}{P_{ss}^{max}}\|^2$;
6: **return** $L_{pred}$.

---

### 2.3.4. Implementation details

In our study, all symbolic regression analyses were performed using PYSR [53], a powerful open-source package developed alongside the Julia library *SymbolicRegression.jl*. The primary model setups are summarized in Table 1. Notably, as mentioned, striking a balance between complexity and expressivity is crucial for the determination of optimal functions in symbolic regression. Figure 2 illustrates an evolutionary process in the search for the target strain energy function ($\Psi(I_2) = 6.3 * \exp(3.5 * [I_2 - 3]^2)$). Synthetic data, incorporating three loading modes of uniaxial tension, uniaxial compression, and simple shear, were generated based on this function, utilizing the mechanical relationships defined in Section 2.3.1. Throughout the evolution process, candidate equations were progressively discovered from scratch, with increasing accuracy and complexity. Ultimately, an equation with sufficient accuracy and acceptable complexity was deemed the optimal strain energy function, closely mirroring the expression of the target function.

In current study, we adopted the default criterion ("*best*") to guide the model selection process. This criterion entails selecting the candidate model with the highest score among expressions and a loss that is at least 1.5 times superior to that of the most accurate model. All training processes took place on a Legion PC equipped with a six-core Intel Core I7-8750H 2.2GHz CPU, 4 GB NVIDIA GTX 1050Ti GPU, and 24GB of memory.

**Table 1. Model setups for symbolic regression analysis.** Exemplary model setups in PYSR for *Invariant-based Symbolic Regression*, *Stretch-based Symbolic Regression*, and *Strain-based Symbolic Regression* algorithms.

| Model setups | Invariant-based SR | Stretch-based SR | Strain-based SR |
|---|---|---|---|
| **Procs** | 6 | 6 | 6 |
| **Populations** | 18 | 18 | 18 |
| **Unary operator** | exp, square, ln_x, cube | None | None |
| **Binary operator** | +, * | +, *, pow_int | +, *, pow_int |
| **Nested_constraints** | ln_x(exp:0, ln_x:0, square:1, cube:1)<br>exp(exp:0, ln_x:0, square:1, cube:1)<br>square(exp:0, ln_x:0, square:0, cube:0)<br>cube(exp:0, ln_x:0, square:0, cube:0) | pow_int(pow_int:0) | pow_int(pow_int:0) |
| **Complexity of operators** | exp:2, ln_x:2, square:3, cube:3, *:1, +:1 | pow_int:2, *:2, +:1 | pow_int:2, *:2, +:1 |
| **Complexity of variables** | 1 | 2 | 2 |
| **Complexity of constants** | 1 | 1 | 1 |
| **Maximum complexity** | 100 | 100 | 100 |
| **Maximum depth** | 10 | 10 | 10 |
| **Niterations** | 1000 | 1000 | 1000 |
| **Enable_autodiff** | True | True | True |
| **User-defined operator** | $\ln\_x(x) = -\ln(1-x)$ | $\text{pow\_int}(x,y) = x^{ceiling(y-0.5)}$ | |

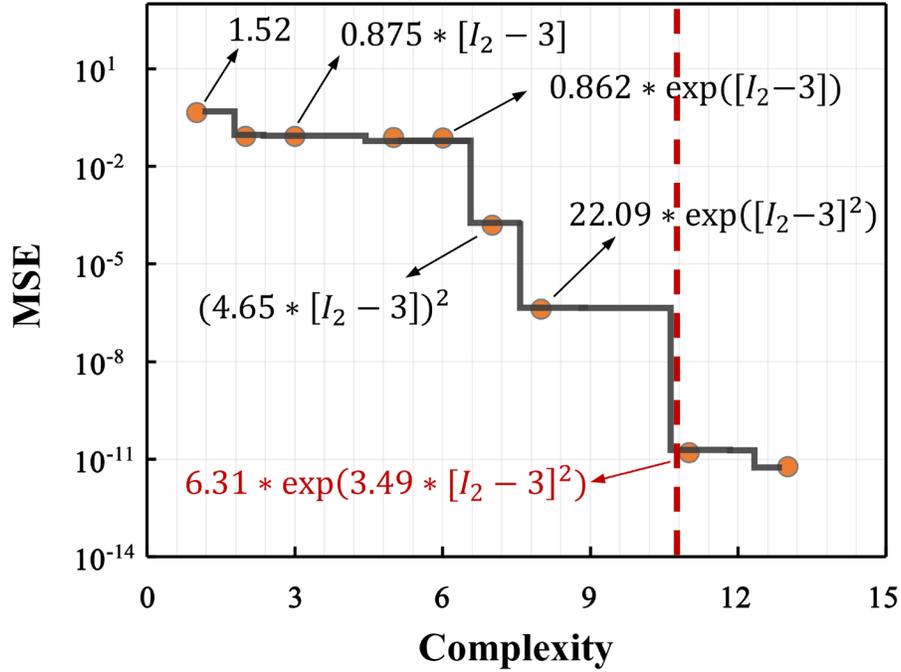

**Figure 2. Optimal function searching process.** Mean square error (MSE) of the discovered strain energy function vs. complexity (parsimony) for synthetic data generated from $\Psi(I_2) = 6.3 * \exp(3.5 * [I_2 - 3]^2)$. Synthetic data consists of three loading modes: uniaxial tension, uniaxial compression, and simple shear. Mean square error is calculated as the sum of each loading mode. Achieving a balance between accuracy and simplicity is essential to determine the optimal functions.

## 3. Numerical Results and Discussions

In this study, our primary objective is to investigate the capability of symbolic regression algorithms in autonomously identifying suitable hyperelastic models for soft tissues, specifically focusing on the human brain cortex. The suitability of a hyperelastic model is characterized by its accuracy, generalizability, and physical admissibility, namely, adherence to the physical constraints, as outlined in Section 2.2. Prior to application, we initially conducted equation search on synthetic dataset to validate the applicability of the algorithms. Subsequently, three distinct endeavors were undertaken for models' discovery on human brain cortex based on multi-mode experimental data, wherein invariants, principal stretches, principal strains were employed as the model inputs, respectively. Finally, we assessed the robustness of symbolic regression by testing the model discovery performance on synthetic data embedded with varying degrees of noise.

## 3.1. Model verification with synthetic data

The synthetic data was generated based on five classical hyperelastic models that are commonly utilized to characterize the material behavior of soft tissues: Mooney Rivlin model, Gent model, Demiray model, Holzapfel model, and Ogden model. The detailed mathematic expressions for each model are as follows:

$$\Psi_{\text{Gent}} = -1.9 \ln(1 - 1.2[I_1 - 3]), \tag{39 a}$$
$$\Psi_{\text{Ogden}} = 0.01(\lambda_1^{-18} + \lambda_2^{-18} + \lambda_3^{-18} - 3), \tag{39 b}$$
$$\Psi_{\text{Demiray}} = 1.66(\exp(0.88[I_1 - 3]) - 1), \tag{39 c}$$
$$\Psi_{\text{Holzapfel}} = 5.6(\exp(3[I_1 - 3]^2) - 1), \tag{39 d}$$
$$\Psi_{\text{Mooney Rivlin}} = 0.87[I_1 - 3] + 0.86[I_1 - 3]^2 + 0.98[I_2 - 3] + 0.43[I_2 - 3]^2. \tag{39 e}$$

Three loading modes (uniaxial tension, uniaxial compression, and simple shear) were considered, with stress data generated in terms of predefined stretches using mechanical expressions defined in section 2.3. The *Invariant-based Symbolic Regression* algorithm was employed for the first four invariant-based models, while the Ogden model data was trained using the *Stretch-based Symbolic Regression* algorithm. The predicted models are visually presented in Figure 3. As shown, all models were successfully predicted using our algorithms, demonstrating precise alignments with the synthetic data.

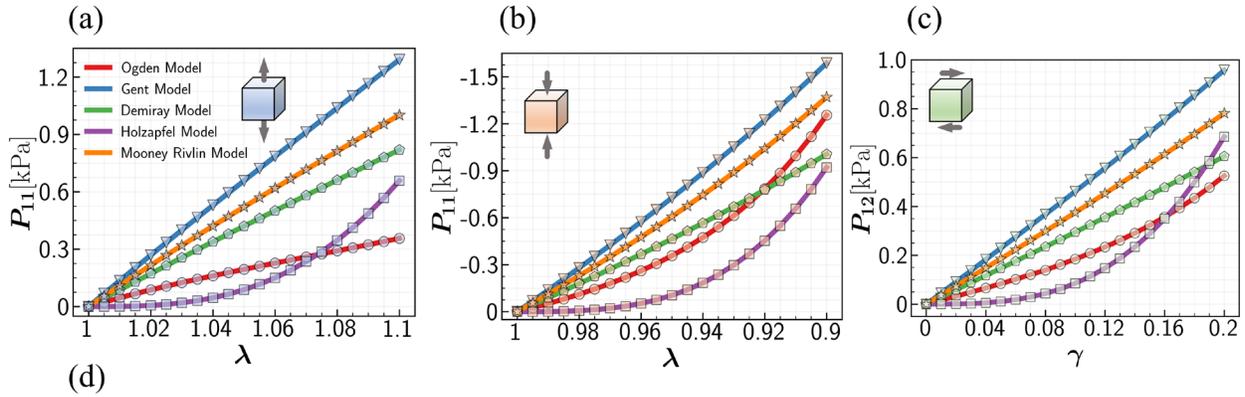

(d)
Ogden model: $\Psi = 0.0102(\lambda_1^{-18} + \lambda_2^{-18} + \lambda_3^{-18} - 3)$    Gent model: $\Psi = -1.899 \ln(1 - 1.201[I_1 - 3])$

Demiray model: $\Psi = 1.661(\exp(0.8802[I_1 - 3]) - 1)$    Holzapfel model: $\Psi = 5.602(\exp(3.001[I_1 - 3]^2) - 1)$

Mooney Rivlin model: $\Psi = 0.8702[I_1 - 3] + 0.8593[I_1 - 3]^2 + 0.9798[I_2 - 3] + 0.4303[I_2 - 3]^2$

**Figure 3. Validation of symbolic regression algorithm**. Models were trained simultaneously with data from three loading modes, and tested with tension (a), compression (b) and shear (c) data individually. Dots illustrate the synthetic data generated from five classical hyperelastic models that widely used for human brain characterization. Specific mathematic expressions of predicted hyperelastic models are provided in (d).

Although the model setups primarily follow the patterns outlined in Table 1, special considerations were made for each algorithm to ensure the accuracy and efficiency in discovering the target functions. First, a constant shift was manually post-defined for Ogden model, Demiray model, and Holzapfel model to ensure the normalization or stress-free condition at the reference state. For example, in the case of the Demiray model with a direct output $\Psi = 1.661 \exp(0.8802[I_1 - 3])$, a constant shift value of -1.661 was prescribed to ensure zero strain energy density ($\Psi = 0$) in the reference configuration ($I_1 = 3$). This constant shift would not affect the stress measure as it may vanish under the differentiation operation. An alternative but automated approach involves customizing the exponential operator as $(\exp(x) - 1)$, similar to the approach used for logarithmic function $(-\ln(1 - x))$. Second, we constrained the order of polynomial operator ($x^y$) to be strictly positive by introducing a ceiling function. Additionally, the value of $y$ was confined within a range of (0, 5) and (-30, 30) for the Mooney Rivlin model and Ogden model, respectively, to enhance efficiency in the equation search. The successful reproductions of these models validate the applicability and accuracy of our algorithms in autonomously discovering the strain energy functions within hyperelastic regimes. Upon validation, we started to perform the symbolic regression based on experimental data, aiming to discover a suitable hyperelastic model for human brain cortex.

### 3.2. Hyperelastic models discovered by *Invariant-based Symbolic Regression* algorithm

Figure 4 illustrates four superior hyperelastic models discovered for the human brain cortex using the *Invariant-based Symbolic Regression* algorithm. These models were trained simultaneously with data from three loading modes, thus the combined loss function defined in Equation (26) was employed for regression optimization, and tested individually for each loading mode. The predictive performance of each model is evaluated by the $R^2$ value, defined as $R^2 = 1 - \sum_{i=1}^{N}(P_i - P_i^*)^2 / \sum_{i=1}^{N}(P_i - \bar{P})^2$, where $\bar{P}$ is the mean of experimental stress. The detailed mathematical expressions of each model are presented at the bottom of the figure. As shown in Figure 4, these four predicted models, though employing distinct formulas, exhibit satisfactory performances in characterizing the material behaviors in uniaxial tension, uniaxial compression, and simple shear scenarios. Notably, all models are solely dependent on the second invariant. This observation aligns with recent findings [33, 54]. Looking at curves, a consistent trend is observed across all models: overestimations in tension occur once the stretch is greater than 5%, underestimation persists throughout compression, and good fitting is observed with simple shear data. The consistent underperformance in uniaxial tension and compression indicates potential data inconsistency with the assumption of hyperelasticity [29]. In essence, the experimental data may not be equally reliable for different loading modes. To address this, we propose introducing weighting factors in Equation (26) to assess the contribution of each loading mode to the combined loss function.

Interestingly, the predicted models $\Psi_a$ and $\Psi_b$ closely resemble the models discovered by constitutive artificial neural network (CANN) regularized with subset selection ($L_0$ regularization) [55]. However, our searched models $\Psi_c$ and $\Psi_d$, demonstrating comparable predictive performance, are not covered in their research findings. Conversely, two hyperelastic models presented in their work with satisfactory fitting accuracy are not discovered by our current algorithm. One is constructed by terms $([I_2 - 3]^2)$ and $(\exp([I_2 - 3]) - 1)$, and another is formed by $(\exp([I_2 - 3]) - 1)$ and $(\exp([I_2 - 3]^2) - 1)$. This observation suggests the potential existence of multiple optima for the current optimization problem. For further exploration, it is crucial to consider either an enriched function space or more diverse loading modes when conducting symbolic regression or CANN. Among the four models we discovered, the third model $\Psi_c = 0.017(\exp(27.91[I_2 - 3]) - 1)$ exhibits the highest predictive accuracy with the simplest form. Therefore, it serves as the optimal model discovered by the *Invariant-based Symbolic Regression* algorithm.

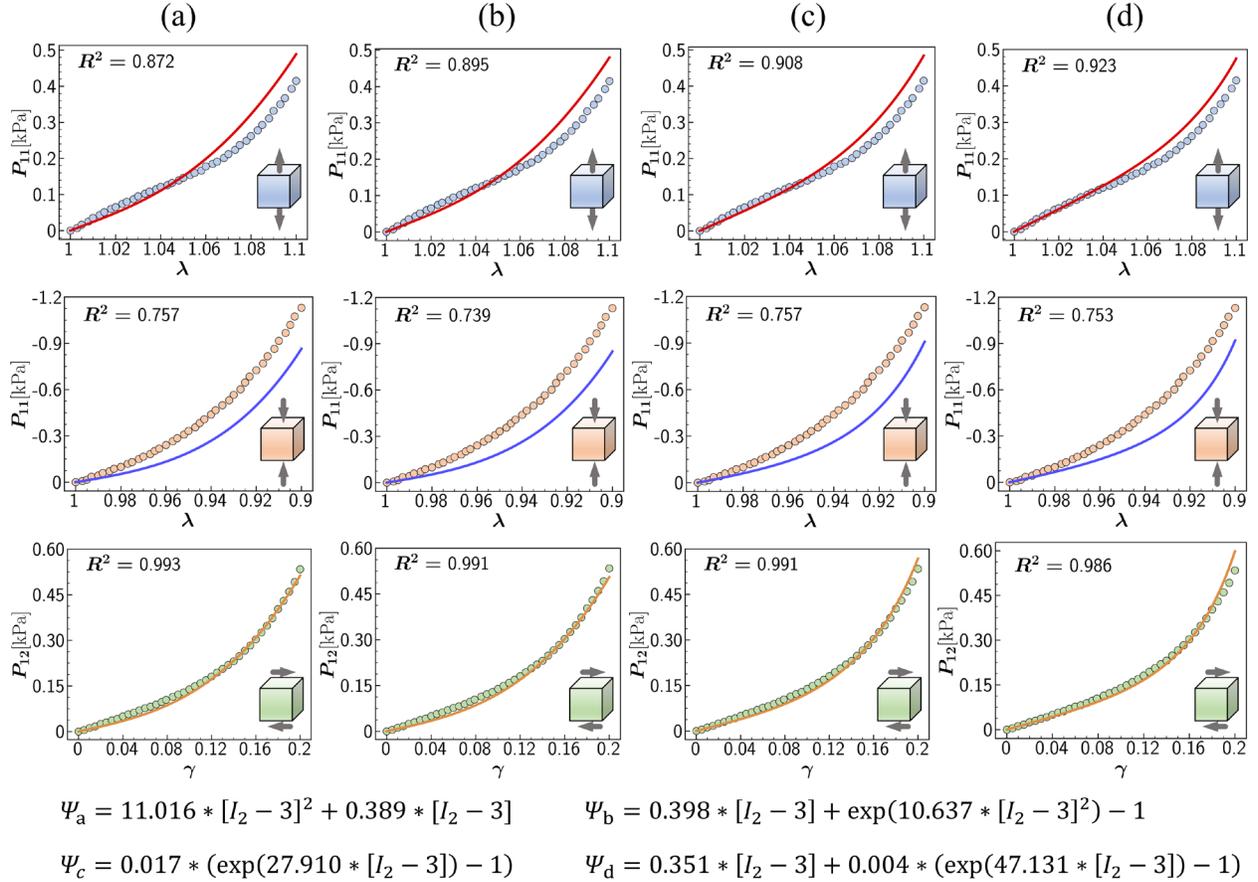

$$\Psi_a = 11.016 * [I_2 - 3]^2 + 0.389 * [I_2 - 3]$$

$$\Psi_b = 0.398 * [I_2 - 3] + \exp(10.637 * [I_2 - 3]^2) - 1$$

$$\Psi_c = 0.017 * (\exp(27.910 * [I_2 - 3]) - 1)$$

$$\Psi_d = 0.351 * [I_2 - 3] + 0.004 * (\exp(47.131 * [I_2 - 3]) - 1)$$

**Figure 4. Four distinct hyperelastic models discovered with invariant-based algorithm.** Models are trained simultaneously with data from three loading modes, and tested with tension, compression and, shear data individually. Dots illustrate the experimental data of the human brain cortex. $R^2$ indicates the goodness of fit. Corresponding mathematical expressions of strain energy function are presented at the bottom of the figure.

Within the framework of invariants, Figure 5 provides a comparison on the predictive performance of the optimal hyperelastic model discovered by symbolic regression, multiple regression, and artificial neural networks. The latter two models are derived from our recent paper [54], where artificial neural networks followed the idea of CANNs [27], but utilized a different loss function, the mean absolute percentage error, MAPE. From the comparison, no significant difference can be observed for these three models, except for a smaller underestimation in tension for the symbolic regression model, which yields a slightly higher $R^2$ value (0.908), as shown in Figure 5a. Notably, the model obtained from multiple regression takes the same format as the symbolic regression model, while the neural network model has a different form, $\Psi_{NN} = 0.552[I_2 - 3] - 2.858 \ln(1 - 2.483[I_2 - 3]^2)$.

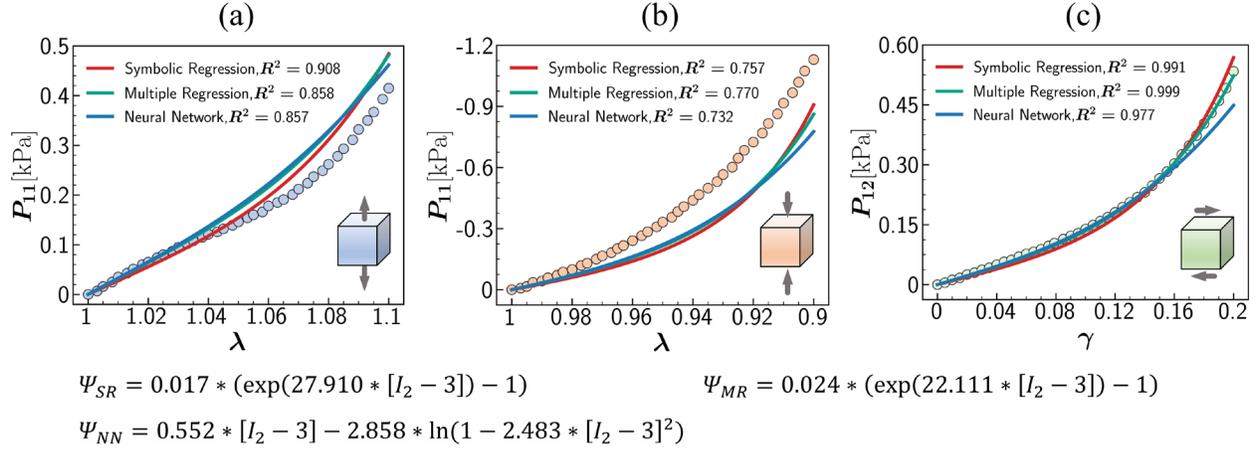

$\Psi_{SR} = 0.017 * (\exp(27.910 * [I_2 - 3]) - 1)$  $\Psi_{MR} = 0.024 * (\exp(22.111 * [I_2 - 3]) - 1)$

$\Psi_{NN} = 0.552 * [I_2 - 3] - 2.858 * \ln(1 - 2.483 * [I_2 - 3]^2)$

**Figure 5. Symbolic regression vs Multiple regression and artificial neural network.** Comparison on predictive performance of invariant-based hyperelastic models derived from symbolic regression (SR), multiple regression (MR), and neural network (NN). Models are trained simultaneously with data from three loading modes, and tested with tension (a), compression (b), and shear data (c), individually. Dots illustrate the experimental data of the human brain cortex. $R^2$ indicates the goodness of fit. Corresponding mathematical expressions of strain energy function are presented at the bottom of the figure.

For invariant-based hyperelastic models, critical physical admissible conditions elucidated in Section 2.2, such as the polyconvexity, have been predefined and embedded within the symbolic regression algorithm. Therefore, the four models unveiled in Figure 4 are inherently designed to satisfy the convexity requirement. To substantiate this claim, we generated 3D surface plots for visualizing the strain energy function with respect to principal stretches $\lambda_1$ and $\lambda_2$, as depicted in Figure 6. The hyperelastic model under consideration has the following expression: $\Psi = 11.016 * [I_2 - 3]^2 + 0.389 * [I_2 - 3]$. Additionally, an incompressibility condition is imposed, resulting in the constraint $\lambda_3 = 1/(\lambda_1 \lambda_2)$. Therefore, the value of $I_2$ can be calculated as: $I_2 = \lambda_1^{-2} + \lambda_2^{-2} + \lambda_1^2 \lambda_2^2$.

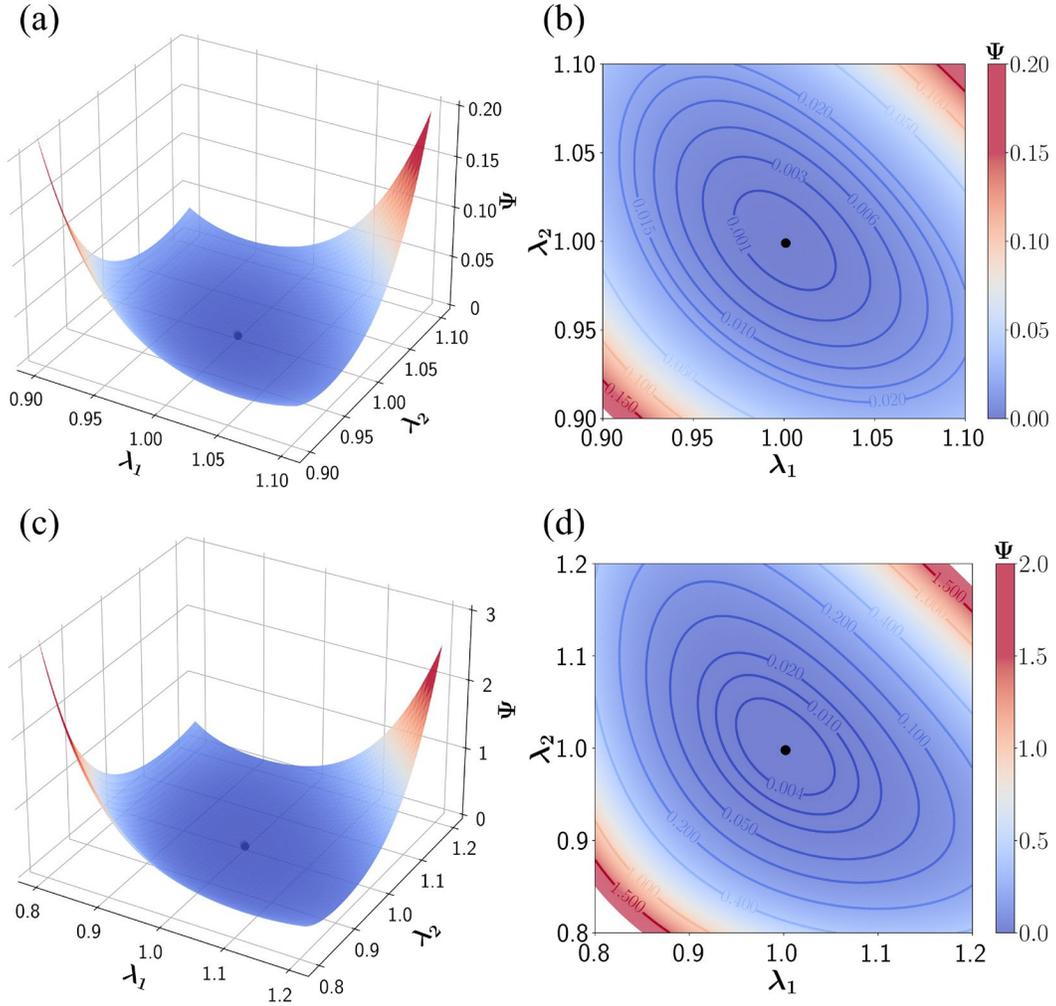

**Figure 6. Convexity of an hyperelastic model derived from invariant-based algorithm**. (a). 3D visualization of the strain energy function $\Psi$ with respect to principal stretch $\lambda_1$ and $\lambda_2$, ($\lambda_3 = 1/(\lambda_1 \lambda_2)$) within the training dataset of human brain cortex; (b). Contour of the strain energy function in (a); (c). 3D visualization of the strain energy density function $\Psi$ with respect to principal stretch $\lambda_1$ and $\lambda_2$ from synthetic data; (d). Contour of the strain energy function in (c). Black dots represent the location of global minimum. The illustrated strain energy function is $\Psi = 11.016 * [I_2 - 3]^2 + 0.389 * [I_2 - 3]$.

In Figure 6a, synthetic stretches within the training dataset of human brain cortex were generated to construct the strain energy function. The location of global minimum is distinctly identified as the dark point. The function demonstrates a pronounced concave shape and attains its minimum at the reference state ($\lambda_1 = 1, \lambda_2 = 1$). Meanwhile, the strain energy function $\Psi$ exhibits significantly high values as both deformations approach their maximum at 10% tension or compression. Further clarity is provided through contour plots in Figure 6b, where higher strain energy densities (warm color) are achieved at the top-right and bottom-left corners. Moreover, the contour lines trace elliptic shapes encircling the center (dark point), thereby signifying the fulfillments of the ellipticity requirement for strain energy functions. The rigorous adherence to polyconvexity condition and coercivity condition robustly establishes the stability, monotony, and uniqueness for the strain energy function under the training dataset [16]. Remarkably, this adherence extends beyond the confines of the training regime. As shown in Figure 6c and 6d, we expanded the synthetic stretches to encompass 20% tensile and compressive deformations. Notably, the strain energy function consistently maintains its convexity, and the contour lines continue to exhibit elliptic shapes. This unwavering consistency attests to the pronounced generalizability of the discovered invariant-based hyperelastic models [33].

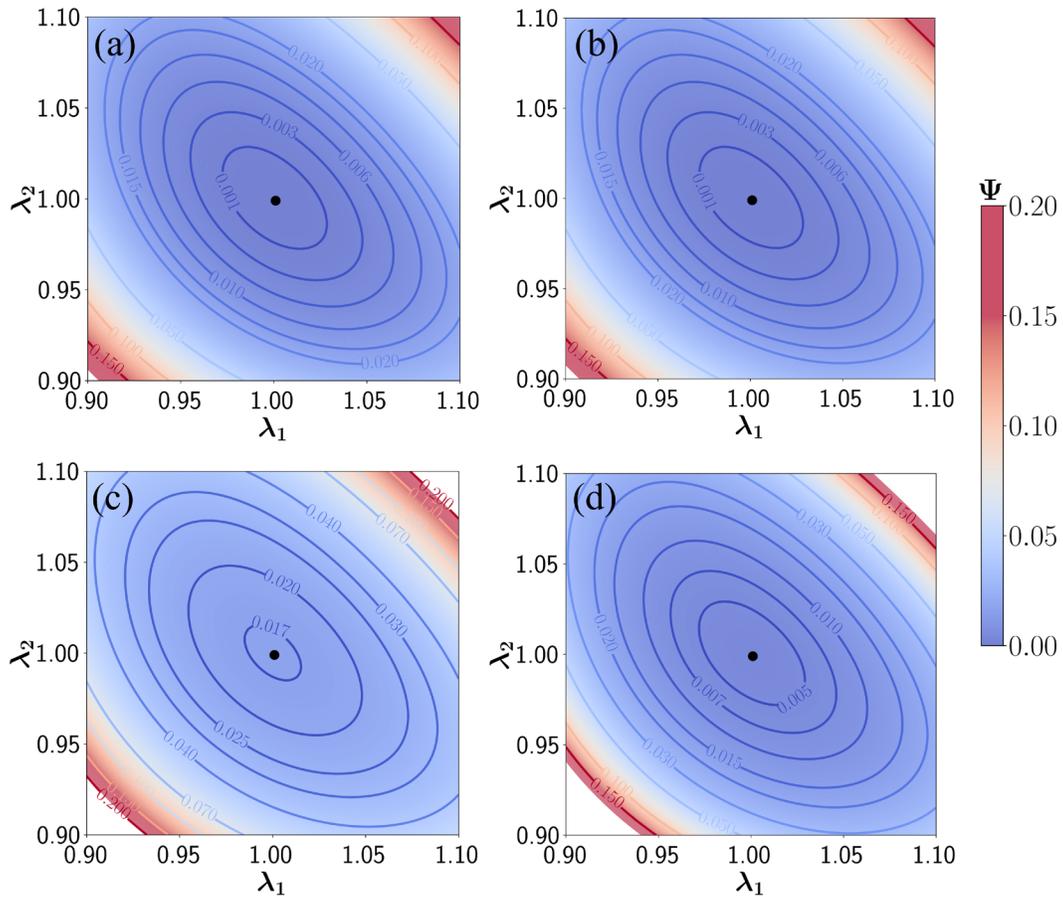

$\Psi_a = 11.016 * [I_2 - 3]^2 + 0.389 * [I_2 - 3]$   $\Psi_b = 0.398 * [I_2 - 3] + \exp(10.637 * [I_2 - 3]^2) - 1$

$\Psi_c = 0.017 * (\exp(27.910 * [I_2 - 3]) - 1)$   $\Psi_d = 0.351 * [I_2 - 3] + 0.004 * (\exp(47.131 * [I_2 - 3]) - 1)$

**Figure 7. Convexity of four invariant-based hyperelastic models within the training data regime.** Contours of the four strain energy functions discovered by *Invariant-based Symbolic Regression* algorithm within the training data regime, (a) $\Psi_a$, (b) $\Psi_b$, (c) $\Psi_c$, (d) $\Psi_d$. Black dots represent the location of global minimum. Corresponding mathematical expressions of strain energy function are presented at the bottom of the figure.

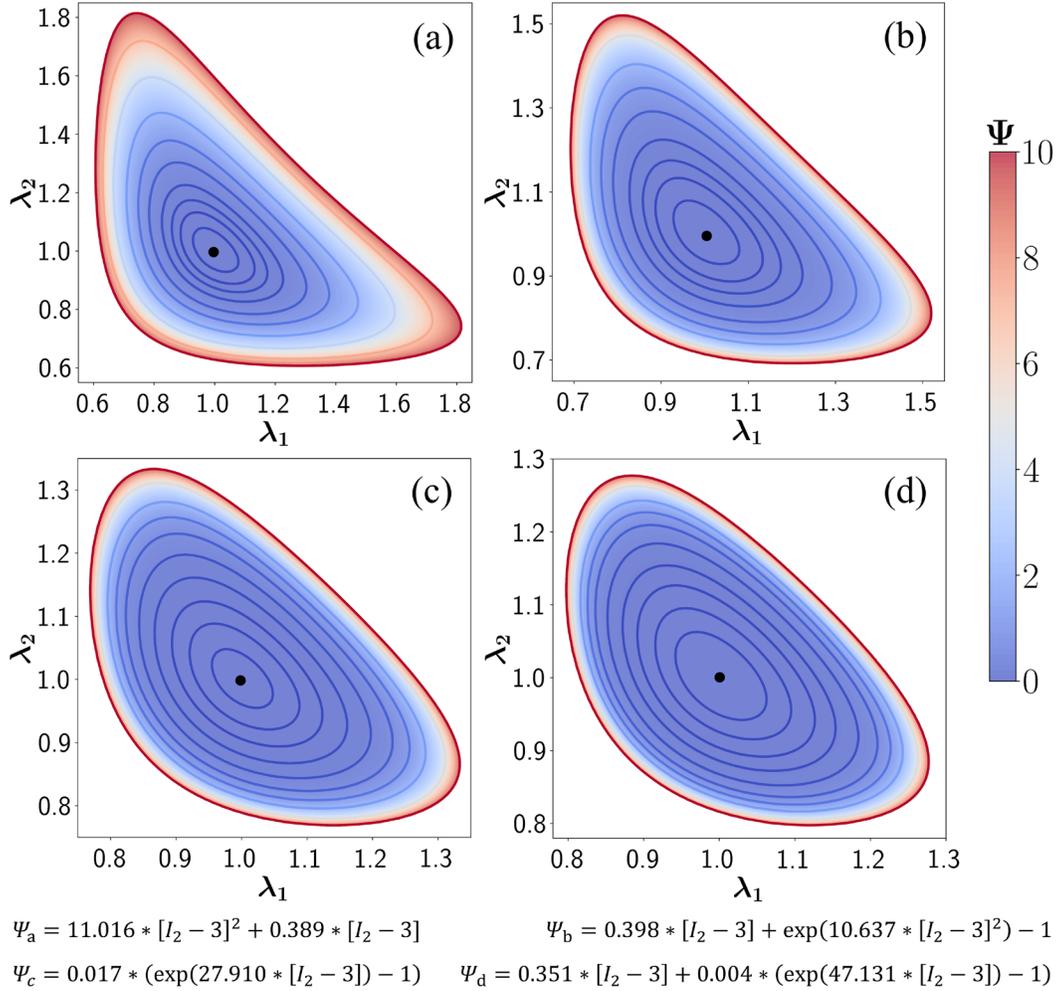

$\Psi_a = 11.016 * [I_2 - 3]^2 + 0.389 * [I_2 - 3]$

$\Psi_b = 0.398 * [I_2 - 3] + \exp(10.637 * [I_2 - 3]^2) - 1$

$\Psi_c = 0.017 * (\exp(27.910 * [I_2 - 3]) - 1)$

$\Psi_d = 0.351 * [I_2 - 3] + 0.004 * (\exp(47.131 * [I_2 - 3]) - 1)$

**Figure 8. Convexity of four invariant-based hyperelastic models beyond the training data regime.** Contours of the four strain energy functions discovered by *Invariant-based Symbolic Regression* algorithm beyond the training data regime, (a) $\Psi_a$, (b) $\Psi_b$, (c) $\Psi_c$, (d) $\Psi_d$. Black dots represent the location of global minimum. Corresponding mathematical expressions of strain energy function are presented at the bottom of the figure.

In addition to the hyperelastic model highlighted in Figure 6, an analysis of the convexity of the remaining three models is presented in Figure 7 and Figure 8. Despite incorporating distinct functional operators, the contour lines of these strain energy functions consistently display elliptic shapes encircling the center point, representing the stress-free reference state ($\lambda_1 = 1, \lambda_2 = 1$), particularly evident in Figure 7. This observation signifies the strict convexity of the strain energy function within the training data regime. Furthermore, the convexity is preserved beyond the training range, as shown in Figure 8. It is noteworthy that the different deformation ranges illustrated in Figure 8 are chosen for enhanced visualization. Ultimately, the physical validity, especially the polyconvexity condition, of all four hyperelastic models discovered by *Invariant-based Symbolic Regression* algorithm have been clearly demonstrated.

### 3.3. Hyperelastic models identified by *Stretch-based Symbolic Regression* algorithm

Figure 9 presents four superior hyperelastic models discovered for the human brain cortex, using the *Stretch-based Symbolic Regression* algorithm. These models are trained simultaneously with data from three loading modes, thus the combined loss function defined in Equation (26) was employed for regression optimization, and tested individually for each mode. The fitting performance of each model is evaluated by the $R^2$ value. The detailed mathematical expressions of each model are provided at the bottom of the figure. All four predicted models exhibit commendable fitting accuracy in characterizing material behaviors within uniaxial tension, uniaxial compression, and simple shear scenarios, boasting $R^2$ value exceeding 0.9 for each loading mode. In contrast to the performance of invariant-based models that exhibit limitations in uniaxial tension and compression, the stretch-based hyperelastic models prove markedly superior in describing uniaxial deformations, particularly compressions. All stretch-based models demonstrate perfect alignment with the compressive data. Conversely, the invariant-based models consistently underestimate compressive forces throughout deformation, leading to $R^2$ values below 0.8, as evident in Figure 4. This observation aligns with recent findings [61]. The superior performance of stretch-based hyperelastic models is particularly pronounced in nonlinear stages, in regions with more than 5% tension or compression, thereby manifesting their efficacy in characterizing the nonlinearity inherent in experimental data [62, 63].

Similar endeavors have been explored using the stretch-based approaches, specifically the generalized Ogden model, to characterize the material behavior of human brain cortex [3, 61]. The one-term Ogden model in [3] shares an identical expression with our first model ($\Psi_a$ in Figure 9). The generalized Ogden models [61] using principal stretch-based CANN exhibit lower accuracy compared to our predicted models ($\Psi_b$, $\Psi_c$, $\Psi_d$), with $R^2$ values of 0.938, 0.985, and 0.987 for tension, compression, and shear, respectively. Notably, the number of terms constituting their model is significantly greater than in ours [61], indicating that we achieved higher accuracy with a more succinct model form. This disparity is attributed to the smaller function space they employed, where the polynomial order is confined within (-30, 10). In contrast, this value is chosen from a wider range of (-30, 30) in our study.

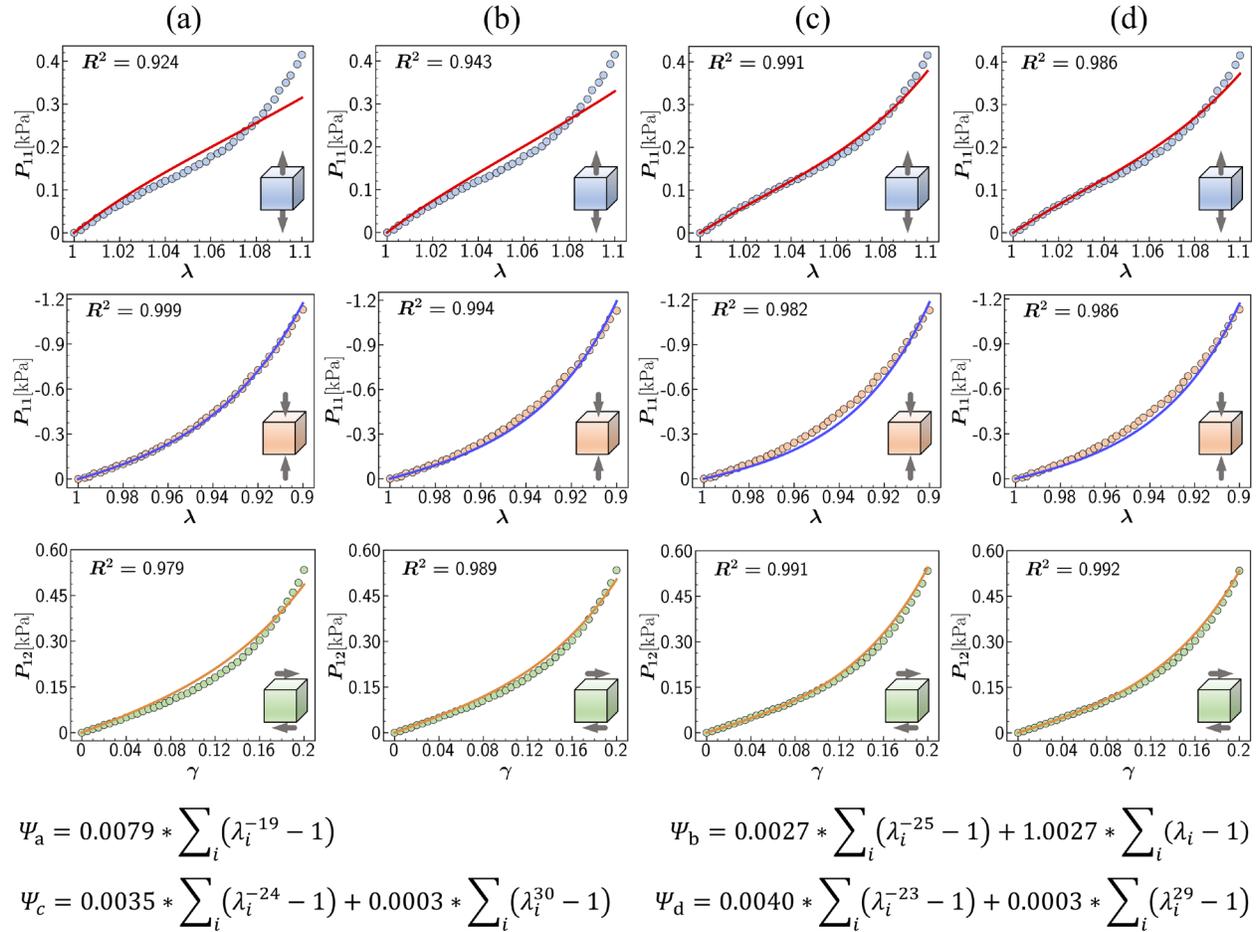

$$\Psi_a = 0.0079 * \sum_i (\lambda_i^{-19} - 1)$$

$$\Psi_b = 0.0027 * \sum_i (\lambda_i^{-25} - 1) + 1.0027 * \sum_i (\lambda_i - 1)$$

$$\Psi_c = 0.0035 * \sum_i (\lambda_i^{-24} - 1) + 0.0003 * \sum_i (\lambda_i^{30} - 1)$$

$$\Psi_d = 0.0040 * \sum_i (\lambda_i^{-23} - 1) + 0.0003 * \sum_i (\lambda_i^{29} - 1)$$

**Figure 9. Four distinct Ogden-form hyperelastic models discovered with stretch-based algorithm**. Models are trained simultaneously with data from three loading modes, and tested with tension, compression, and shear data individually. Dots illustrate the experimental data of the human brain cortex. $R^2$ indicates the goodness of fit. Corresponding mathematical forms of strain energy density function are presented at the bottom of the figure.

Among the four models presented in Figure 9, the last two models, $\Psi_c = 0.0035 \sum_i (\lambda_i^{-24} - 1) + 0.0003 \sum_i (\lambda_i^{30} - 1)$, $\Psi_d = 0.004 \sum_i (\lambda_i^{-23} - 1) + 0.0003 \sum_i (\lambda_i^{29} - 1)$, demonstrate comparably high accuracy in fitting the multi-mode data. As such, these models serve as the optimal outcomes derived from the *Stretch-based Symbolic Regression* algorithm. The utilization of large exponents in these equations further validates the high nonlinearity inherent in the experimental data, implying the inherent intricacy present in the behavior of human brain cortex [64].

Unlike the invariant-based hyperelastic models that strictly enforce the output function to be polyconvex, stretch-based may exhibit non-convexity due to the random combinations of polynomial series with orders in arbitrary values and signs [65]. Hence, we conducted convexity checks for each model derived from *Stretch-based Symbolic Regression* algorithm by assessing whether the Hessian matrix (second derivatives of $\Psi$ with respect to principal stretch $\lambda_i$), as defined in Equation (16), is positive definite. The results are presented in Table 2. The positive determinants and real eigenvalues of the Hessian matrix ensure the positive definiteness of tangent ($\partial^2 \Psi / \partial \lambda_i^2$), further confirming the convexity of strain energy functions within the experimental dataset. A more straightforward representation is shown in Figure 9, where we generated 3D surface plots for illustrating the strain energy function with respect to principal stretches $\lambda_1$ and $\lambda_2$. The hyperelastic model under consideration has the following expression: $\Psi = 0.0035 * \sum_i (\lambda_i^{-24} - 1) + 0.0003 * \sum_i (\lambda_i^{30} - 1)$. Additionally, an incompressibility condition is imposed, resulting in the constraint $\lambda_3 = 1/(\lambda_1 \lambda_2)$.

**Table 2. Convexity checks for stretch-based hyperelastic models**. $\partial^2 \Psi / \partial \lambda_i^2$ is the second derivatives of $\Psi$ with respect to principal stretch $\lambda_i$, with $i$ in 1, 2, 3; $\min(\det[H])$ is the minimal determinant of Hessian matrix $[H]$; $\min(\partial^2 \Psi / \partial \lambda_i^2)$ denotes the minimal value of the $i$th eigenvalue of $[H]$. $\Psi_a$, $\Psi_b$, $\Psi_c$, $\Psi_d$ correspond to the four stretch-based hyperelastic models shown in Figure 9. Experimental data of human brain cortex are employed for calculations.

| Convexity Checks | $\Psi_a$ | $\Psi_b$ | $\Psi_c$ | $\Psi_d$ |
|---|---|---|---|---|
| $\partial^2\Psi/\partial\lambda_i^2$ | $2.998\lambda_i^{-21}$ | $1.751\lambda_i^{-27}$ | $0.274\lambda_i^{28} + 2.081\lambda_i^{-26}$ | $0.260\lambda_i^{27} + 2.2051\lambda_i^{-25}$ |
| min (det[$H$]) | 26.951 | 5.369 | 13.074 | 14.983 |
| min($\partial^2\Psi/\partial\lambda_1^2$) | 0.405 | 0.134 | 1.569 | 1.577 |
| min($\partial^2\Psi/\partial\lambda_2^2$) | 0.405 | 0.134 | 1.569 | 1.577 |
| min($\partial^2\Psi/\partial\lambda_3^2$) | 0.036 | 0.006 | 1.568 | 1.576 |
| ? Positive definite | Yes | Yes | Yes | Yes |
| ? Convexity of $\Psi$ | **Yes** | **Yes** | **Yes** | **Yes** |

In Figure 10a, we generated synthetic stretches within the training range to construct the strain energy function. The location of global minimum is identified as the dark point. The strain energy function exhibits a noticeable concave shape with its minimum located at the center ($\lambda_1 = 1, \lambda_2 = 1$). Meanwhile, the strain energy function $\Psi$ tends to reach its maximum under 10% tensions or compressions in both directions. A quantitative illustration is presented in the contour plot in Figure 10b, where the concavity is evident through radiatively increased iso-energy lines. These lines form near-elliptic shapes encircling the global minimum, indicating the fulfillment of the ellipticity requirement for strain energy functions. Furthermore, we observe a notable diagonal symmetry in these counter lines, bisecting the two axes and contributing to the isotropy of strain energy functions. Similar symmetry patterns are present for the invariant-based strain energy functions, as depicted in Figure 7. This symmetry is primarily attributed to the incompressibility constraint imposed to the invariants and principal stretches. Within the incompressible framework, all discovered invariant-based models are reliant on the second invariant, $I_2 = \lambda_1^{-2} + \lambda_2^{-2} + \lambda_1^2\lambda_2^2$, which is equally dependent on $\lambda_1$ and $\lambda_2$, indicating that the interchange of these two principal stretches will not alter the value of $I_2$. Analogously, for stretch-based models such as the first model represented in Figure 9, $\Psi_a = 0.0079 * \sum_i(\lambda_i^{-19} - 1) = 0.0079 * (\lambda_1^{-19} + \lambda_2^{-19} + \lambda_3^{-19} - 3)$. The incompressibility condition ($\lambda_3 = 1/(\lambda_1\lambda_2)$) yields $\Psi_a = 0.0079 * (\lambda_1^{-19} + \lambda_2^{-19} + \lambda_1^{19}\lambda_2^{19} - 3)$, leading to the equivalence between $\lambda_1$ and $\lambda_2$.

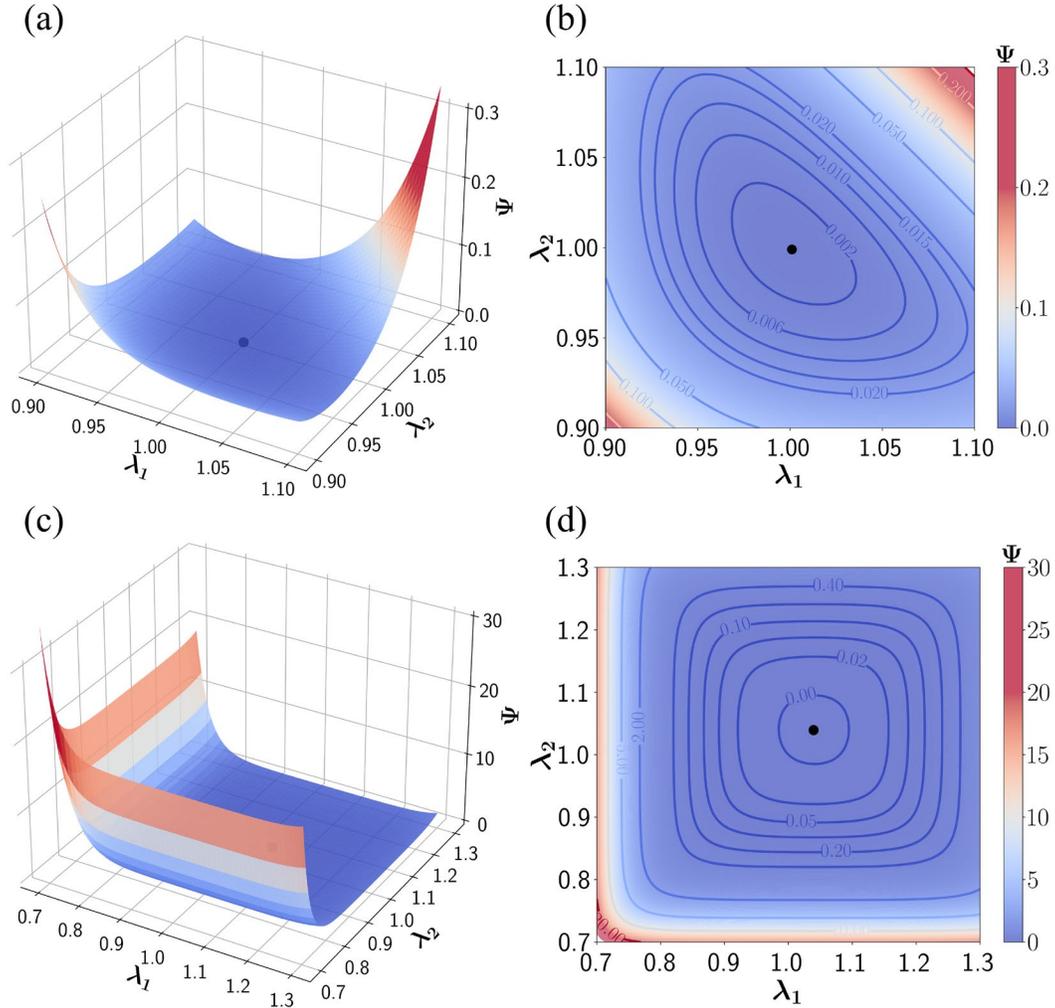

**Figure 10. Convexity of an hyperelastic model derived from stretch-based algorithm**. (a). 3D visualization of the strain energy density function $\Psi$ with respect to principal stretch $\lambda_1$ and $\lambda_2$, ($\lambda_3 = 1/(\lambda_1\lambda_2)$) within the training dataset of human brain cortex; (b). Contour of the strain energy function in (a); (c). 3D visualization of the strain energy density function $\Psi$ with respect to principal stretch $\lambda_1$ and $\lambda_2$, ($\lambda_3 = 1$) from synthetic data; (d). Contour of the strain energy function in (c). Black dots represent the location of global minimum. The strain energy function employed for illustration is $\Psi = 0.0035 * \sum_i(\lambda_i^{-24} - 1) + 0.0003 * \sum_i(\lambda_i^{30} - 1)$.

Continuing this line of thought, we attempted to relax the incompressibility constraint and assume no deformation occurring in the third principal direction ($\lambda_3 = 1$), corresponding to a plane strain state. The resultant 3D surface and contour of the strain energy function are shown in Figure 10c and 10d, respectively. As expected, the contribution of $\lambda_1$ and $\lambda_2$ to the strain energy function is decoupled, resulting in mutually perpendicular contour lines in two principal directions.

Notably, the global minimum (around $\lambda_1 = \lambda_2 = 1.03$) for this case shifts from the original stress-free state, and the coercivity condition might be violated as the strain energy function converges to a finite value when both principal stretches reach their maximum. This suggests that additional adjustments, i.e., incorporating boundary conditions, are necessary for ensuring stability and uniqueness. Nevertheless, the noted imbalance in the stretch-based strain energy functions during tension ($\lambda_i > 1$) and compression ($\lambda_i < 1$) further underscore the effectiveness of the generalized Ogden model in addressing tension-compression nonlinearity inherent in soft tissues [66].

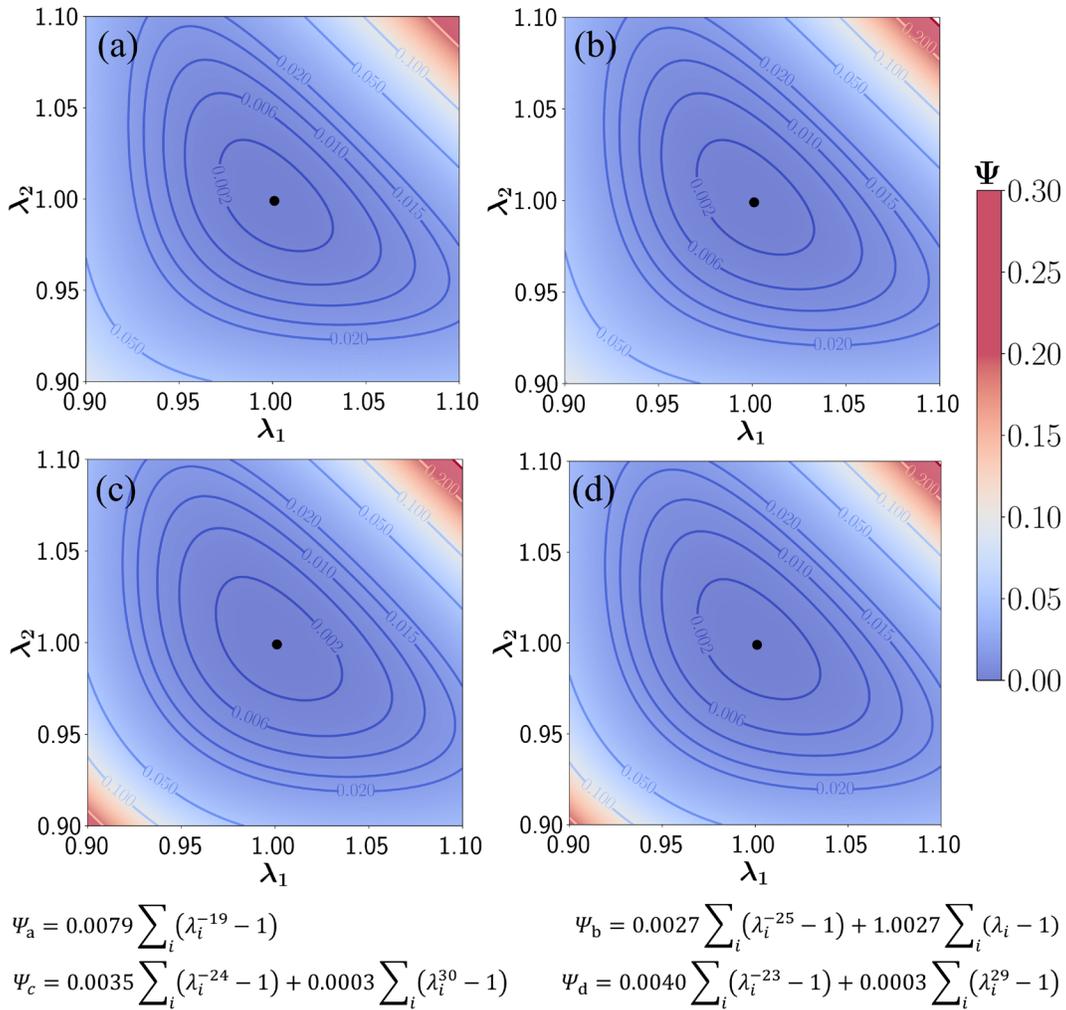

$\Psi_a = 0.0079 \sum_i (\lambda_i^{-19} - 1)$

$\Psi_b = 0.0027 \sum_i (\lambda_i^{-25} - 1) + 1.0027 \sum_i (\lambda_i - 1)$

$\Psi_c = 0.0035 \sum_i (\lambda_i^{-24} - 1) + 0.0003 \sum_i (\lambda_i^{30} - 1)$

$\Psi_d = 0.0040 \sum_i (\lambda_i^{-23} - 1) + 0.0003 \sum_i (\lambda_i^{29} - 1)$

**Figure 11. Convexity of four stretch-based hyperelastic models within the training data regime.** Contours of the four strain energy functions discovered by *Stretch-based Symbolic Regression* algorithm within the training data regime, (a) $\Psi_a$, (b) $\Psi_b$, (c) $\Psi_c$, (d) $\Psi_d$. Black dots

represent the location of global minimum. Corresponding mathematical expressions of strain energy function are presented at the bottom of the figure.

Representations of the convexity of all four stretch-based hyperelastic models are present in Figure 11 and Figure 12. In Figure 11, the contour lines all exhibit near-elliptic shapes encircling the center points that correspond to the stress-free reference state ($\lambda_1 = 1, \lambda_2 = 1$). This observation confirms that the predicted strain energy functions satisfy the polyconvexity or ellipticity requirement within the training data regime. Furthermore, we expanded the synthetic stretches ranges to examine whether the convexity is preserved beyond the training range. Results are shown in Figure 12, where different deformation ranges are selected for enhanced visualization. Though all models present concave shapes with global minimum at the center and higher value at the outer boundaries, the strict adherence to convexity requirements is not maintained, particularly for $\Psi_a$ in Figure 12a and $\Psi_b$ in Figure 12b. In Figure 12a, the contour lines resemble a right triangle, with the hypotenuse posing a significant non-convex shape. The issue is also evident for $\Psi_b$ in Figure 12b, suggesting a potential loss of convexity for $\Psi_a$ and $\Psi_b$ under large deformations. Similar observations were also reported in investigations of the one-term Ogden model [65]. Nonetheless, the remaining two models ($\Psi_c$ and $\Psi_d$) still preserve their convexity beyond the training data regime, as illustrated in Figure 12c and 12d. Overall, the convexity of hyperelastic models discovered by *Stretch-based Symbolic Regression* algorithm is clearly demonstrated, though caution is warranted for potential convexity loss in specific scenarios, as observed in $\Psi_a$ and $\Psi_b$ under large deformations.

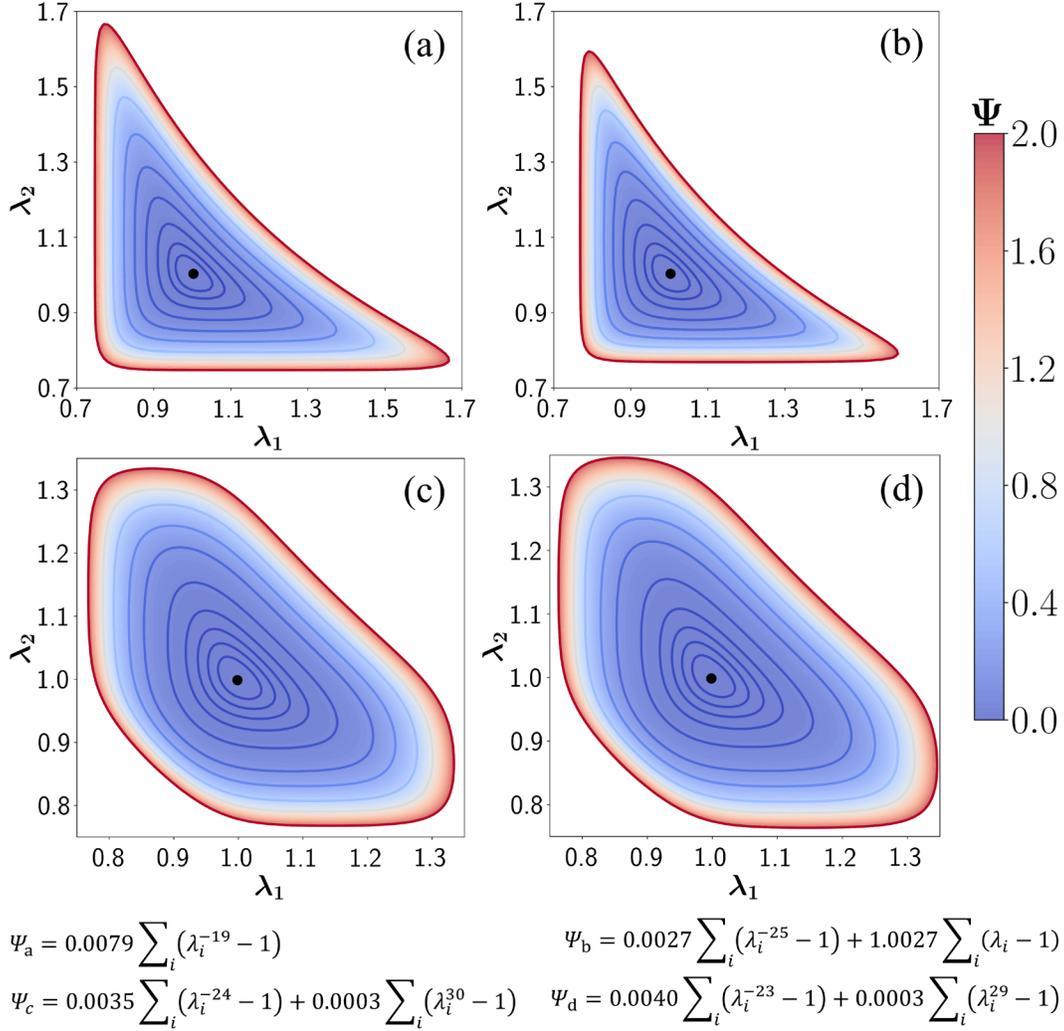

$$\Psi_a = 0.0079 \sum_i (\lambda_i^{-19} - 1)$$

$$\Psi_b = 0.0027 \sum_i (\lambda_i^{-25} - 1) + 1.0027 \sum_i (\lambda_i - 1)$$

$$\Psi_c = 0.0035 \sum_i (\lambda_i^{-24} - 1) + 0.0003 \sum_i (\lambda_i^{30} - 1)$$

$$\Psi_d = 0.0040 \sum_i (\lambda_i^{-23} - 1) + 0.0003 \sum_i (\lambda_i^{29} - 1)$$

**Figure 12. Convexity of four stretch-based hyperelastic models beyond the training data regime.** Contours of the four strain energy functions discovered by *Stretch-based Symbolic Regression* algorithm beyond the training data regime, (a) $\Psi_a$, (b) $\Psi_b$, (c) $\Psi_c$, (d) $\Psi_d$. Black dots represent the location of global minimum. Corresponding mathematical expressions of strain energy function are presented at the bottom of the figure.

### 3.4. Hyperelastic models predicted by *Strain-based Symbolic Regression* algorithm

Unlike the *Invariant-based Symbolic Regression* or principal *Stretch-based Symbolic Regression* algorithms, the *Strain-based Symbolic Regression* algorithm predicted only one unique model,

$$\Psi = \sum_i (2820.76 \epsilon_i^6 + 43.27 \epsilon_i^4 - 13.72 \epsilon_i^3 + 1.37 \epsilon_i^2). \tag{40}$$

Here, $\epsilon_i$ represents the normal strain (also known as Biot strain) and is related to principal stretch by $\epsilon_i = \lambda_i - 1$. The predictive performance of this strain-based model is shown in Figure 13, where the goodness of fit for each model is evaluated by the $R^2$ value. As shown, the strain-based model demonstrates exceptional fitting accuracy in capturing material behaviors within uniaxial tension, uniaxial compression, and simple shear scenarios, with $R^2$ values exceeding 0.99.

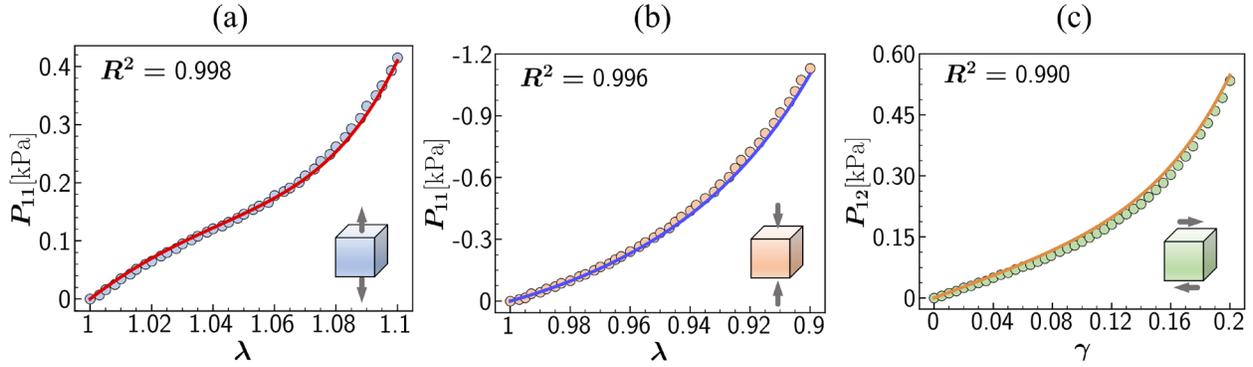

**Figure 13. Hyperelastic model discovered with strain-based algorithm**. Models are trained simultaneously with data from three loading modes, and tested with tension (a), compression (b) and, shear (c) data individually. Dots illustrate the experimental data of the human brain cortex. $R^2$ indicates the goodness of fit.

Figure 14 depicts a comparison among the optimal models predicted by *Invariant-based Symbolic Regression*, *Stretch-based Symbolic Regression*, and *Strain-based Symbolic Regression* algorithms, respectively. As seen, the strain-based model exhibits the highest fitting accuracy with experimental data across all three loading modes. This remarkable fitting behavior pertaining to the strain-based model was also reported in our recently published paper [56], where the multiple regression method, combined with the Akaike's information criteria (AIC), was employed to identify the optimal hyperelastic model within a confined polynomial space with an order's range of (0,10). Interestingly, the polynomial series set selected by multiple regression model is identical to our current symbolic regression model ($\epsilon_i^6$, $\epsilon_i^4$, $\epsilon_i^3$, $\epsilon_i^2$), while the coefficients differ. Additionally, the polynomial orders available for symbolic regression algorithm are confined within (-30, 30), indicating that the polynomial set discovered by multiple regression is indeed the optimal one. It is noteworthy that though multiple regression demonstrates satisfactory capability in model discovery, substantial human efforts and computational costs are required, especially when dealing with a vast functional space. For example, when the polynomial space is confined within (-30, 30), the potential equation combinations amount to $2^{60}$, making it impractical to traverse completely using multiple regression. This underscores the advantages of evolution algorithms over traditional regression methods like multiple regression in functions searching and parameter optimization [67, 68].

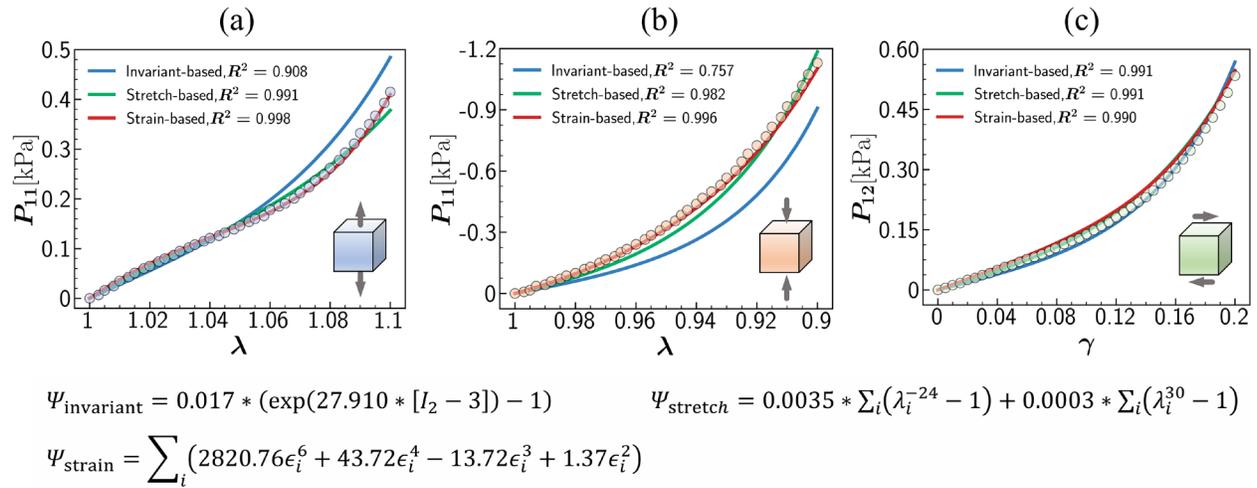

$$\Psi_{\text{invariant}} = 0.017 * (\exp(27.910 * [I_2 - 3]) - 1)$$

$$\Psi_{\text{stretch}} = 0.0035 * \sum_i (\lambda_i^{-24} - 1) + 0.0003 * \sum_i (\lambda_i^{30} - 1)$$

$$\Psi_{\text{strain}} = \sum_i (2820.76\epsilon_i^6 + 43.72\epsilon_i^4 - 13.72\epsilon_i^3 + 1.37\epsilon_i^2)$$

**Figure 14. Symbolic regression with invariant-based algorithm vs stretch-based algorithm and strain-based algorithm.** Comparison on predictive performance of optimal hyperelastic models derived from symbolic regression using three distinct algorithms. Models are trained simultaneously with data from three loading modes, and tested with tension (a), compression (b), and shear (c) data individually. Dots illustrate the experimental data of the human brain cortex. $R^2$ indicates the goodness of fit. Corresponding mathematical expressions of strain energy functions are presented at the bottom of the figure.

Analogously, the convexity of the strain-based model was determined by evaluating the positive-definiteness of the Hessian matrix, as defined in Equation (16). Within the training data regime, the minimal determinant of the Hessian matrix is 3.158, and the minimal value for the three eigenvalues ($\partial^2\Psi/\partial\lambda_1^2$, $\partial^2\Psi/\partial\lambda_2^2$, $\partial^2\Psi/\partial\lambda_3^2$) are 0.422, 0.422, and 0.420, respectively. The positive determinant and real eigenvalues ensure the positive definiteness of the Hessian matrix, which further confirms the convexity of the strain energy function within the training dataset. Furthermore, the consistency condition (positivity of shear modulus), as defined in

Equation (38), is also guaranteed, as the coefficient before the second order term ($\epsilon_i^2$) is positive (1.37).

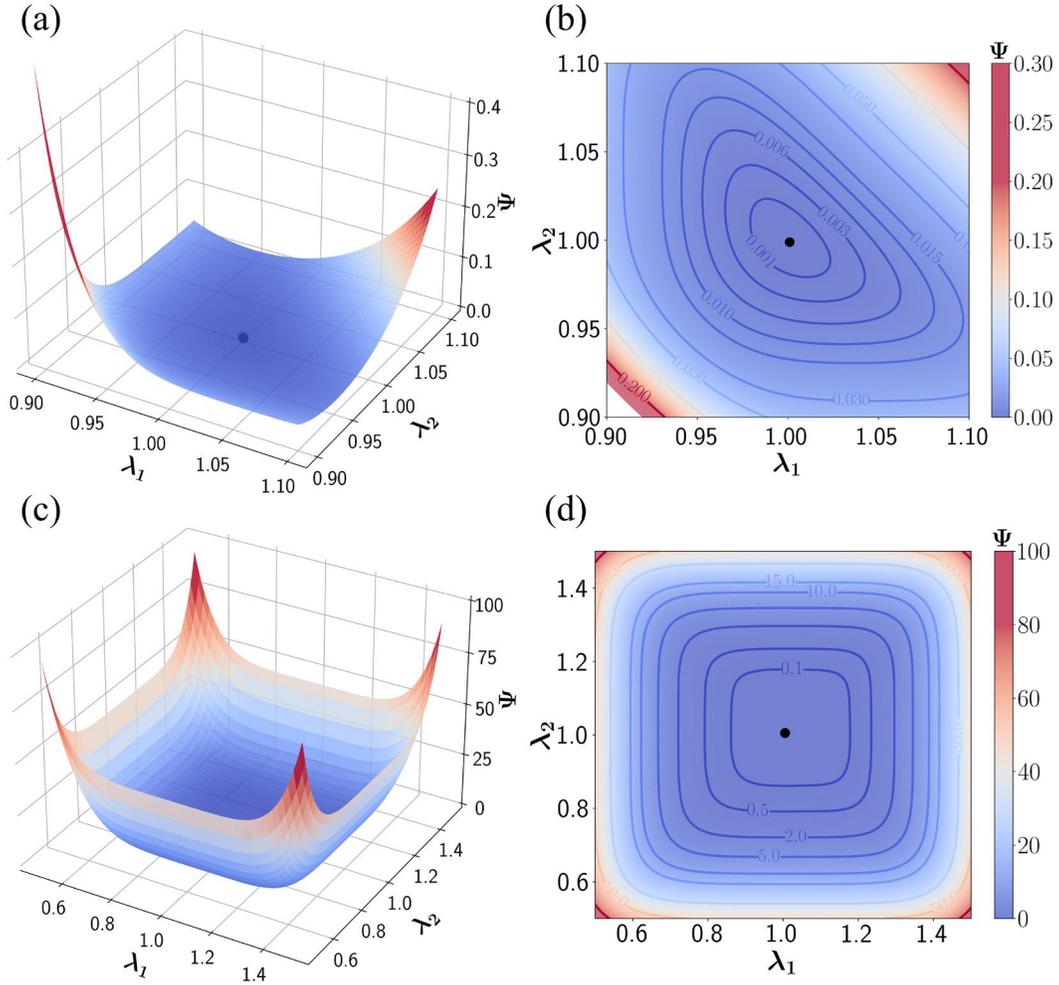

**Figure 15. Convexity of the hyperelastic model derived from strain-based algorithm**. (a). 3D visualization of the strain energy density function $\Psi$ with respect to principal stretch $\lambda_1$ and $\lambda_2$, ($\lambda_3 = 1/(\lambda_1\lambda_2)$) within the training dataset of human brain cortex; (b). Contour of the strain energy function in (a); (c). 3D visualization of the strain energy density function $\Psi$ with respect to principal stretch $\lambda_1$ and $\lambda_2$, ($\lambda_3 = 1$) from synthetic data; (d). Contour of the strain energy function in (c). Black dots represent the location of global minimum.

A more straightforward representation is provided by the 3D surface plots and contour plots, as illustrated in Figure 15. Within the incompressibility framework (Figure 15a and 15b), the strain energy function exhibits a pronounced concave shape, and its minimum is attained at the reference state ($\lambda_1 = \lambda_2 = 1$). Additionally, the coercivity condition is also satisfied, as the strain energy function continues to achieve its local maximum with increasing tensile or compressive deformations. However, the peak strain energy function under compression is significantly higher than its tensile counterpart under the same deformation. The distinct behaviors under tension and compression potentially explains the prominence of strain-based models in capturing the data nonlinearity, as shown in Figure 13. If we relax the compressibility constraint, the strain energy demonstrates perfect symmetry with respect to $\lambda_1$ and $\lambda_2$, as seen in Figure 15c and 15d.

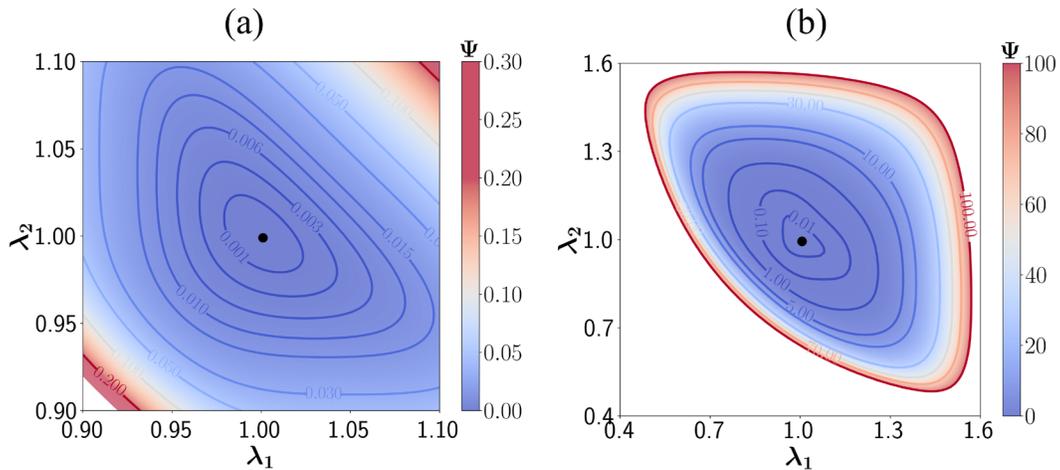

**Figure 16. Convexity of the strain-based hyperelastic model within and beyond the training data regime.** Contours of the strain energy function discovered by *Strain-based Symbolic Regression* algorithm within (a) and beyond (b) the training data regime. Black dots represent the location of global minimum.

To examine the convexity beyond the training data regime, we expanded the synthetic stretches ranges to 60% tension and compression and created a contour plot, as presented in Figure 16. Notably, the incompressibility constraint is imposed in this situation. From the figure, convexity observed in the training regime (Figure 16a) is preserved beyond the training dataset (Figure 16b), even in the case with 60% deformation. This result is intriguing considering we applied loose restrictions to strain energy functions in the *Strain-based Symbolic Regression* algorithm, where both the coefficients and polynomial orders are not constrained to be strictly positive, leading to the negative coefficient observed in Equation (40). Moreover, the strain-based functions $f(\epsilon_i)$ can be expanded into the stretch-based polynomials $g(\lambda_i)$ using the relations $\epsilon_i = \lambda_i - 1$, further indicating the stricter confinements prescribed for stretch-based energy function. However, the strain-based energy function still exhibits rigorous ellipticity, and convexity is well preserved even in large deformations, under which the stretch-based energy functions may lose their convexity, see Figure 12. This unanticipated result suggests that the constraints we applied to enforce the polyconvexity or positivity of shear modulus are potentially over-restrictive, and a loose prior confinement coupled with rigorous posterior checks may generate more satisfactory results.

### 3.5. Robustness investigation for symbolic regression algorithms

In this section, we investigated the robustness of our symbolic regression algorithms employed for hyperelastic model discovery. The baseline data originated from the optimal model discovered by the *Invariant-based Symbolic Regression* algorithm, namely, $\Psi = 0.0170 * (\exp(27.9100 * [I_2 - 3]) - 1)$. Artificial Gaussian noises were incorporated into the synthetic data to mimic the perturbations encountered in real experiments,

$$P_{i,k}^{\text{test}} = P_{i,k}^{\text{synthetic}} + P_{i,k}^{\text{noise}}, \quad P_{i,k}^{\text{noise}} \sim \mathcal{N}(0, \sigma_k) \quad \forall\, i \in \{1, \ldots, n_{data}\}, \quad k \in \{\text{ut}, \text{uc}, \text{ss}\}, \quad (41)$$

where $k$ represents loading modes: uniaxial tension (ut), uniaxial compression (uc), and simple shear (ss); $P_{i,k}^{\text{synthetic}}$ means the $i$th synthetic stress data under $k$th loading mode; $P_{i,k}^{\text{noise}}$ denotes the prescribed noise, sampled from a Gaussian distribution with zero mean and standard deviation $\sigma_k$. Here, we applied a consistent relative deviation to each loading mode, thus the actual standard deviation was scaled based on the maximal stress in each mode, $\sigma_k = \sigma * P_{\max,k}$. In this study, relative deviations ranging from 0 to 20% were utilized to assess the robustness of our symbolic regression algorithms. The model setup for each noise case remained consistent, as outlined in Table 1.

**Table 3. Robustness tests**. Effects of prescribed noise on symbolic regression predictions. The target strain energy function is represented as $\Psi = 0.0170 * (\exp(27.9100 * [I_2 - 3]) - 1)$. A prediction is deemed right when the mathematic format of the prediction model coincides with the target model.

| Noise, $\sigma$ | MSE | Predicted $\Psi$ | ? Right Prediction |
|---|---|---|---|
| 0 | $1.50 \times 10^{-11}$ | $0.0170 * (\exp(27.9100 * [I_2 - 3]) - 1)$ | Yes |
| 0.0001 | $2.20 \times 10^{-10}$ | $0.0170 * (\exp(27.9085 * [I_2 - 3]) - 1)$ | Yes |
| 0.001 | $1.02 \times 10^{-8}$ | $0.0170 * (\exp(27.9063 * [I_2 - 3]) - 1)$ | Yes |
| 0.002 | $1.51 \times 10^{-6}$ | $0.0171 * (\exp(27.8814 * [I_2 - 3]) - 1)$ | Yes |
| 0.005 | $1.15 \times 10^{-5}$ | $0.0170 * (\exp(27.8227 * [I_2 - 3]) - 1)$ | Yes |
| 0.01 | $3.99 \times 10^{-5}$ | $0.0170 * (\exp(27.8616 * [I_2 - 3]) - 1)$ | Yes |
| 0.02 | $2.04 \times 10^{-4}$ | $0.0168 * (\exp(28.0941 * [I_2 - 3]) - 1)$ | Yes |
| 0.05 | $1.05 \times 10^{-3}$ | $0.0168 * (\exp(28.1500 * [I_2 - 3]) - 1)$ | Yes |
| 0.1 | $4.79 \times 10^{-3}$ | $0.0227 * (\exp(23.4639 * [I_2 - 3]) - 1)$ | Yes |
| 0.2 | $2.10 \times 10^{-2}$ | $[I_2 - 3]$ | No |

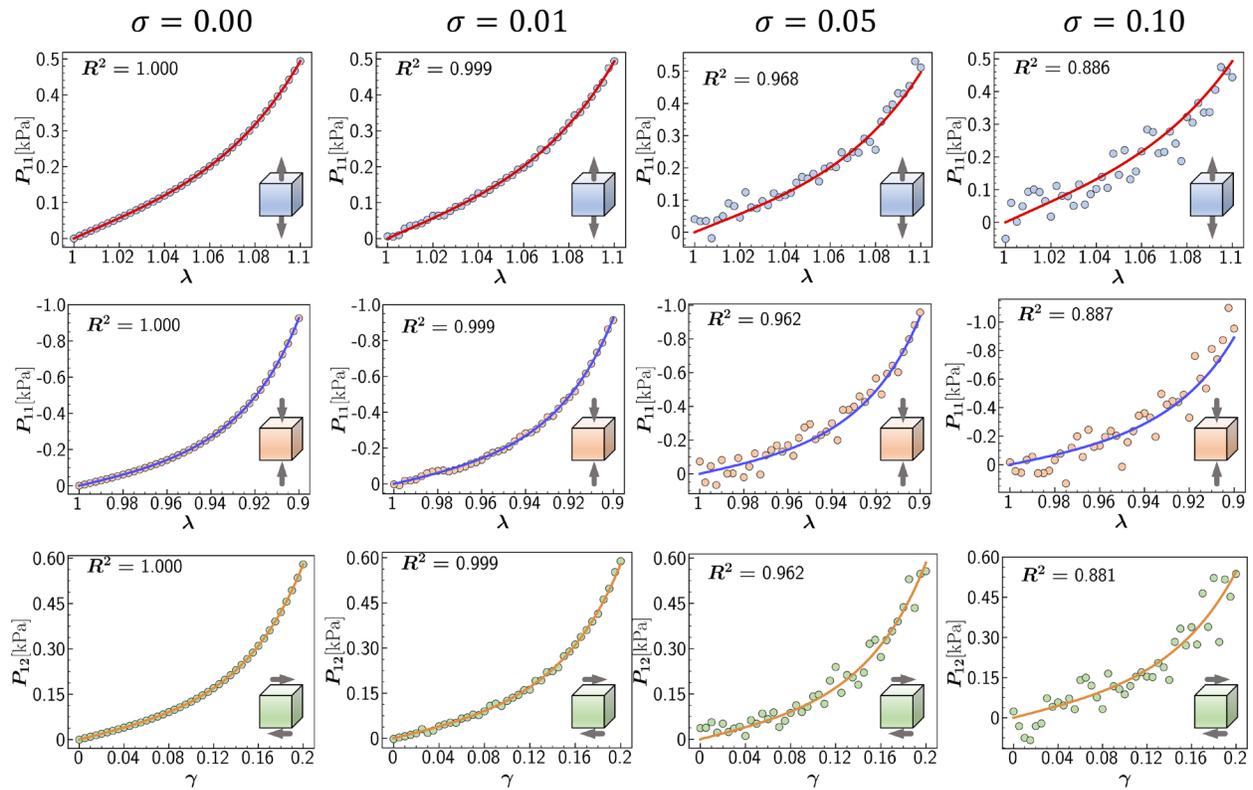

**Figure 17. Robustness test.** Effects of prescribed noise on the predictions of symbolic regression. Models are trained simultaneously with synthetic data from three loading modes, and tested with tension, compression, and shear data individually. Dots illustrate the experimental data of the human brain cortex. $R^2$ indicates the goodness of fit.

The prediction results along with the corresponding MSE are shown in Table 3, while the predictive performances under four typical noise scenarios are illustrated in Figure 17. As seen, our algorithm is capable of precisely discovering the predefined mathematic format of the target strain energy function even with a 10% noise prescription. However, the accuracy of data fitting continues to diminish as noise amplifies. With a 20% noise imposition, the data distributions become too random to a discernible mathematic trend, resulting in the failure of precise prediction. Nevertheless, it is still conclusive that our algorithms demonstrate robustness in current model discovery scenarios.

## 4. Conclusion and Future Endeavors

We proposed a symbolic regression framework to autonomously discover interpretable hyperelastic models from sparse experimental data, while adhering to physical laws. To ensure the physical validity of the predicted constitutive models, we designed the symbolic regression algorithm, especially the objective functions, by integrating various physical constraints such as the polyconvexity condition. We explored three distinct hyperelastic models, invariant-based, principal stretch-based, and normal strain-based, to unveil the capabilities of our symbolic regression algorithms. After validating on synthetic data, we extended our study to the human brain cortex, using experimental data across three loading modes. We demonstrated convexity within and beyond the experimental data for each discovered model. Finally, we assessed the robustness of our algorithms using synthetic data embedded with artificial Gaussian noises.

Our results reveal that symbolic regression is capable of discovering accurate models with parsimonious mathematic expressions for invariant-based, stretch-based, and strain-based cases. Among all discovered models, the strain-based model exhibits superior performance in fitting the experimental data, with accuracy measure $R^2$ value exceeding 0.99 for all loading modes. Additionally, principal stretch-based and strain-based models effectively captured the nonlinearity and tension-compression asymmetry inherent in the human brain cortex. Polyconvexity checks validate the rigorous fulfillment of convexity within the training data regime and satisfactory extrapolation capabilities beyond the training dataset for all models. Nonetheless, the stretch-based hyperelastic models may lose convexity under large deformations in certain scenarios. Robustness tests underscore the accuracy and precision of our proposed symbolic regression algorithms.

In present study, we leveraged symbolic regression for identifying constitutive material models for the human brain cortex within the hyperelasticity context. Naturally, our approach is readily applicable to model discovery for other brain regions, including the corona radiate and corpus callosum [3], or other soft tissues like skin [69, 70] and muscles [71]. In addition to hyperelasticity, the exploration of other constitutive behaviors in soft tissues, such as viscosity and plasticity, presents intriguing future avenues, akin to similar investigations in alloy composites [41] and concrete beams [51]. Furthermore, our current model discovery utilized data from three loading modes, uniaxial tension, uniaxial compression, and simple shear. Incorporating a more diverse range of loading scenarios, such as biaxial experiments [72], will significantly contribute to the comprehensive characterization of material behaviors, particularly in complex loading cases. It is worth noting that the inclusion of diverse loading data results in a high-dimensional dataset, and the integration of symbolic regression and neural networks may prove beneficial for model discovery in these complex scenarios [73]. Moreover, the framework can also incorporate with Finite Element models to perform inverse parameter identification [74].

**Acknowledgement**

JH and XW acknowledges the support from National Science Foundation (IIS-2011369) and National Institutes of Health (1R01NS135574-01). EK acknowledges the partial support from National Science Foundation (CMMI-2320933).

**Data Availability Statement**

The original contributions presented in the study are included in the article/supplemental material. Further inquiries can be directed to the corresponding authors.

**Conflict of Interest**

The authors declare that the research was conducted in the absence of any commercial or financial relationships that could be construed as a potential conflict of interest.

**Authorship Contribution Statement**

JH: Methodology, Software, Validation, Investigation, Writing – Original Draft; XC: Formal Analysis, Writing – Review and Editing; TW: Validation, Writing – Review and Editing; EK: Validation, Writing – Review and Editing; XW: Conceptualization, Validation, Supervision, Funding acquisition, Writing – Original Draft, Writing – Review and Editing.